\documentclass{aastex63}
 \usepackage{amsmath}
  \usepackage{graphicx}
  \usepackage{subfigure}
  \usepackage{comment}
  \usepackage{bm}
  \usepackage{amssymb}
  \usepackage{scalerel}
  \usepackage{esvect}
 \usepackage{epstopdf}  

\newcommand{\kepler}{{\em Kepler}}
\newcommand{\corot}{{\em CoRoT}}
\newcommand{\numax}{\mbox{$\nu_{\rm max}$}}
\newcommand{\Dnu}{\mbox{$\Delta \nu$}}

\newcommand{\muHz}{\mbox{$\mu$Hz}}

\newcommand{\Dpi}{\Delta \Pi}

\newcommand{\msun}{\!{M_{\sun}}}

\shorttitle{Asteroseismic analysis of HD~222076}
\shortauthors{Jiang et al.}

\begin{document}

\title{\em{TESS} Asteroseismic Analysis of the Known Exoplanet Host Star HD~222076}

\correspondingauthor{Chen Jiang}
\email{jiangch53@mail.sysu.edu.cn}

\author[0000-0002-7614-1665]{Chen Jiang}
\affiliation{School of Physics and Astronomy, Sun Yat-Sen University, No. 135, Xingang Xi Road, Guangzhou, 510275, P. R. China}

\author[0000-0001-5222-4661]{Timothy R. Bedding}
\affiliation{Stellar Astrophysics Centre (SAC), Department of Physics and Astronomy, Aarhus University, Ny Munkegade 120, 8000 Aarhus C, Denmark}
\affiliation{Sydney Institute for Astronomy (SIfA), School of Physics, University of Sydney, NSW 2006, Australia}

\author[0000-0002-3481-9052]{Keivan G. Stassun}
\affiliation{Vanderbilt University, Department of Physics and Astronomy, 6301 Stevenson Center Ln., Nashville, TN 37235, USA}
\affiliation{Vanderbilt Initiative in Data-intensive Astrophysics (VIDA), 6301 Stevenson Center Ln., Nashville, TN 37235, USA}

\author[0000-0001-8014-6162]{Dimitri Veras}
\altaffiliation{STFC Ernest Rutherford Fellow}
\affiliation{Centre for Exoplanets and Habitability, University of Warwick, Coventry CV4 7AL, UK}
\affiliation{Department of Physics, University of Warwick, Coventry CV4 7AL, UK}

\author[0000-0001-8835-2075]{Enrico Corsaro}
\affiliation{INAF --- Osservatorio Astrofisico di Catania, via S.~Sofia 78, 95123 Catania, Italy}

\author[0000-0002-1988-143X]{Derek L. Buzasi}
\affiliation{Department of Chemistry \& Physics, Florida Gulf Coast University, 10501 FGCU Boulevard S., Fort Myers, FL 33965, USA}

\author[0000-0001-8916-8050]{Przemys\l{}aw Miko\l{}ajczyk}
\affiliation{Astronomical Institute, University of Wroc\l{}aw, Miko\l{}aja Kopernika 11, 51-622 Wroc\l{}aw, Poland}

\author[0000-0003-2449-6226]{Qian-sheng, Zhang}
\affiliation{Yunnan Observatories, Chinese Academy of Sciences, 396 Yangfangwang, Guandu District, Kunming, 650216, People's Republic of China}
\affiliation{Center for Astronomical Mega-Science, Chinese Academy of Sciences, 20A Datun Road, Chaoyang District, Beijing, 100012, People's Republic of China}
\affiliation{Key Laboratory for the Structure and Evolution of Celestial Objects, Chinese Academy of Sciences, 396 Yangfangwang, Guandu District, Kunming, 650216, People's Republic of China}
\affiliation{University of Chinese Academy of Sciences, Beijing, 100049, People's Republic of China}

\author[0000-0002-6176-7745]{Jian-wen, Ou}
\affiliation{School of Physics and Astronomy, Sun Yat-Sen University, No. 135, Xingang Xi Road, Guangzhou, 510275, P. R. China}

\author[0000-0002-4588-5389]{Tiago L. Campante}
\affiliation{Instituto de Astrof\'{\i}sica e Ci\^{e}ncias do Espa\c{c}o, Universidade do Porto,  Rua das Estrelas, 4150-762 Porto, Portugal}
\affiliation{Departamento de F\'{\i}sica e Astronomia, Faculdade de Ci\^{e}ncias da Universidade do Porto, Rua do Campo Alegre, s/n, 4169-007 Porto, Portugal}

\author[0000-0002-9414-339X]{Tha\'{\i}se S. Rodrigues}
\affiliation{Osservatorio Astronomico di Padova—INAF, Vicolo dellOsservatorio 5, I-35122 Padova, Italy}

\author[0000-0002-4647-2068]{Benard Nsamba}
\affiliation{Max-Planck-Institut f\"{u}r Astrophysik, Karl-Schwarzschild-Str. 1, D-85748 Garching, Germany}
\affiliation{Instituto de Astrof\'{\i}sica e Ci\^{e}ncias do Espa\c{c}o, Universidade do Porto,  Rua das Estrelas, 4150-762 Porto, Portugal}

\author[0000-0002-9480-8400]{Diego Bossini}
\affiliation{Instituto de Astrof\'{\i}sica e Ci\^{e}ncias do Espa\c{c}o, Universidade do Porto,  Rua das Estrelas, 4150-762 Porto, Portugal}

\author[0000-0002-7084-0529]{Stephen R. Kane}
\affiliation{Department of Earth and Planetary Sciences, University of California, Riverside, CA 92521, USA}

\author[0000-0001-7664-648X]{Jia Mian Joel Ong}
\affiliation{Department of Astronomy, Yale University, P.O.~Box 208101, New Haven, CT 06520-8101, USA}

\author[0000-0002-7772-7641]{Mutlu Y{\i}ld{\i}z}
\affiliation{Department of Astronomy and Space Sciences, Science Faculty, Ege
University, 35100, Bornova, \.Izmir, Turkey.}

\author[0000-0002-9424-2339]{Zeynep \c{C}e\.{i}ik Orhan}
\affiliation{Department of Astronomy and Space Sciences, Science Faculty, Ege
University, 35100, Bornova, \.Izmir, Turkey.}

\author[0000-0001-5759-7790]{S\.{i}bel \"Ortel}
\affiliation{Department of Astronomy and Space Sciences, Science Faculty, Ege
University, 35100, Bornova, \.Izmir, Turkey.}

\author[0000-0001-6832-4325]{Tao Wu}
\affiliation{Yunnan Observatories, Chinese Academy of Sciences, 396 Yangfangwang, Guandu District, Kunming, 650216, People's Republic of China}
\affiliation{Key Laboratory for the Structure and Evolution of Celestial Objects, Chinese Academy of Sciences, 396 Yangfangwang, Guandu District, Kunming, 650216, People's Republic of China}
\affiliation{Center for Astronomical Mega-Science, Chinese Academy of Sciences, 20A Datun Road, Chaoyang District, Beijing, 100012, People's Republic of China}

\author[0000-0003-1860-1851]{Xinyi Zhang}
\affiliation{Yunnan Observatories, Chinese Academy of Sciences, 396 Yangfangwang, Guandu District, Kunming, 650216, People's Republic of China}
\affiliation{Key Laboratory for the Structure and Evolution of Celestial Objects, Chinese Academy of Sciences, 396 Yangfangwang, Guandu District, Kunming, 650216, People's Republic of China}
\affiliation{University of Chinese Academy of Sciences, Beijing, 100049, People's Republic of China}

\author[0000-0001-6396-2563]{Tanda Li}
\affiliation{Stellar Astrophysics Centre (SAC), Department of Physics and Astronomy, Aarhus University, Ny Munkegade 120, 8000 Aarhus C, Denmark}
\affiliation{Sydney Institute for Astronomy (SIfA), School of Physics, University of Sydney, NSW 2006, Australia}
\affiliation{School of Physics and Astronomy, University of Birmingham, Birmingham B15 2TT, UK}

\author[0000-0002-6163-3472]{Sarbani Basu}
\affiliation{Department of Astronomy, Yale University, P.O.~Box 208101, New Haven, CT 06520-8101, USA}

\author[0000-0001-8237-7343]{Margarida S. Cunha}
\affiliation{Instituto de Astrof\'{\i}sica e Ci\^{e}ncias do Espa\c{c}o, Universidade do Porto,  Rua das Estrelas, 4150-762 Porto, Portugal}
\affiliation{Departamento de F\'{\i}sica e Astronomia, Faculdade de Ci\^{e}ncias da Universidade do Porto, Rua do Campo Alegre, s/n, 4169-007 Porto, Portugal}
\affiliation{School of Physics and Astronomy, University of Birmingham, Birmingham B15 2TT, UK}

\author[0000-0001-5137-0966]{J\o rgen Christensen-Dalsgaard}
\affiliation{Stellar Astrophysics Centre (SAC), Department of Physics and Astronomy, Aarhus University, Ny Munkegade 120, 8000 Aarhus C, Denmark}

\author[0000-0002-5714-8618]{William J. Chaplin}
\affiliation{School of Physics and Astronomy, University of Birmingham, Birmingham B15 2TT, UK}
\affiliation{Stellar Astrophysics Centre (SAC), Department of Physics and Astronomy, Aarhus University, Ny Munkegade 120, 8000 Aarhus C, Denmark}

\begin{abstract}

The \textit{Transiting Exoplanet Survey Satellite (TESS)} is an all-sky survey mission aiming to search for exoplanets that transit bright stars. The high-quality photometric data of \textit{TESS} are excellent for the asteroseismic study of solar-like stars. In this work, we present an asteroseismic analysis of the red-giant star HD~222076 hosting a long-period (2.4 yr) giant planet discovered through radial velocities. Solar-like oscillations of HD~222076 are detected around  $203 \, \muHz$ by \textit{TESS} for the first time. Asteroseismic modeling, using global asteroseismic parameters as input, yields a determination of the stellar mass ($M_\star =  1.12 \pm 0.12\, \msun$), radius ($R_\star = 4.34 \pm 0.21\,R_\sun$), and age ($7.4 \pm 2.7\,$Gyr), with precisions greatly improved from previous studies. The period spacing of the dipolar mixed modes extracted from the observed power spectrum reveals that the star is on the red-giant branch burning hydrogen in a shell surrounding the core.  We find that the planet will not escape the tidal pull of the star and be engulfed into it within about $800\,$Myr, before the tip of the red-giant branch is reached.
\end{abstract}


\section{Introduction} \label{sec:intro}

Thanks to the very long duration and high precision of photometric space observation missions, such as \corot\ \citep{baglin06} and \kepler\ \citep{borucki10}, asteroseismology has made major advances in the understanding of stellar interior physics and evolution. In particular, the detection of oscillations in solar-type and red-giant stars has led to breakthroughs such as the discovery of fast core rotation and a way to distinguish between hydrogen-shell-burning stars and stars that are also burning helium in their cores \citep[see][for a review]{chaplin13}.  

With the development of data analysis techniques \citep[e.g.,][]{davies16, lund17a, corsaro14,corsaro15} and improved stellar modeling strategies \citep[e.g.,][]{seren17,silva17}, as well as optimization procedures that make use of individual oscillation frequencies \citep{metcalfe10, jiang11, mathur12, silva15, randle19}, asteroseismology has also proven to be an efficient tool to estimate fundamental stellar properties such as stellar masses, radii and ages. This, in turn, enables the systematic characterization of exoplanet host stars through asteroseismology, which provides with unmatched precision in the absolute properties of their planets \citep{ballard14,camp15,silva15,lundkvist16,camp19,huber19}. Furthermore, the synergy between exoplanet research and asteroseismology also enables us to set constraints on the spin-orbit alignment of exoplanet systems \citep{huber13, benomar14, chaplin14a, lund14, camp16a, kamiaka19} and to perform statistical inferences on the orbital eccentricities, through asterodensity profiling \citep{kane12,sliski14,van15,van19}.

The NASA‘s \textit{Transiting Exoplanet Survey Satellite (TESS)} Mission \citep{ricker14} is performing a near all-sky survey for exoplanets using the transit method in an area 400 times larger than that covered by the \kepler\ mission, reinforcing the synergy between asteroseismology and exoplanet science. Using its dedicated 2-minute cadence and excellent photometric precision observations, \textit{TESS} is expected to detect oscillations in thousands of solar-like oscillators \citep{camp16b, schofield19}, and simulations predict that at least 100 solar-type and red-giant stars observed by \textit{TESS} will host transiting or nontransiting exoplanets \citep{camp16b}. Considering the geometric transit probability of each system detected by the radial velocity (RV) method and the observational strategy of \textit{TESS}, \cite{dalba19} predicted that \textit{TESS} would observe transits for $\sim$11 RV-detected planets in its primary mission. However, only three of these detections were expected to be novel, such that the RV-detected planet was not previously known to transit.

In this paper, we present an asteroseismic analysis of the evolved known host HD~222076, which has a long-period planet detected through RV method. \textit{TESS} observed solar-like oscillations for HD~222076 for the first time. We use these oscillations in detailed stellar modeling to derive the mass, radius and age of the host star. Detections of oscillations by \textit{TESS} in the previously known exoplanet-host stars HD~212771 and HD~203949 were reported by \cite{camp19}, following on the discovery of the first planet transiting a star (TOI-197 or TESS Object of Interest 197) in which oscillations could be measured \citep{huber19}.

\section{Observations}

\subsection{High-Resolution Spectroscopy}
\label{sc:spectro}

HD~222076 (TIC~325178933, HIP~116630) is a bright (with apparent \textit{TESS} magnitude $T = 6.59$), spectroscopically-classified red-giant-branch star \citep[K1 III;][]{houk75}. It is among the targets of an RV survey of 164 bright G and K giant stars in the southern hemisphere conducted by \cite{jones11}, with the purpose of studying the effect of the host star evolution on the inner structure of planetary systems.
And it is also listed in the Stars With ExoplanETs CATalog \citep[SWEET-Cat,][]{santos13, sousa18}, which provides stellar atmospheric parameters and masses for the planet-host stars derived assuming local thermodynamic equilibrium (LTE) and using high-resolution and high signal-to-noise spectra. Based on precise radial velocities obtained with three instruments in two parallel planet-search efforts: the UCLES spectrograph \citep{diego90} on the 3.9\,m Anglo-Australian Telescope, the CHIRON spectrograph \citep{toko13} on the 1.5\,m telescope at CTIO, and the FEROS spectrograph on the 2.2\,m telescope at La Silla \citep{kaufer99}, \cite{witt17} reported the detection of a giant planet HD~222076b that has an orbital period of $P = 871 \pm 19\,$days with a semimajor axis of $a = 1.83 \pm 0.03\,$au, and a minimum mass of $m \sin i = 1.56 \pm 0.11\, M_{\rm Jup}$ (with $i$ being the orbital inclination and $M_{\rm Jup}$ the mass of Jupiter).
A complete list of stellar parameters and relevant literature sources are given in Table~\ref{tb:starpar}. 


\subsection{Broadband Photometry and \textnormal{Gaia} Parallax}
\label{sc:sed}

\begin{figure}
\resizebox{1.0\hsize}{!}{\includegraphics[angle=90]{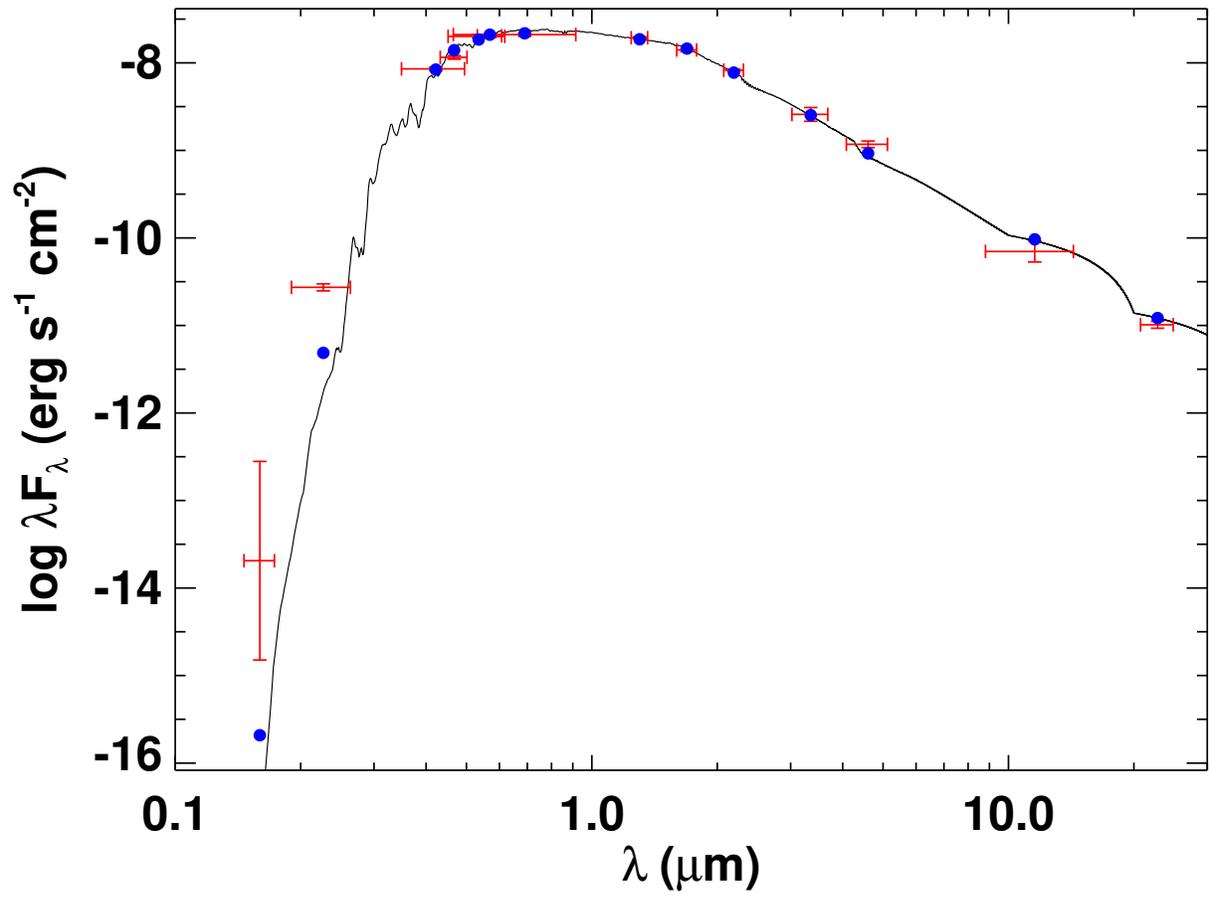}}
\caption{Spectral energy distribution (SED). Red symbols represent the observed photometric measurements, where the horizontal bars represent the effective width of the passband. Blue symbols are the model fluxes from the best-fit Kurucz atmosphere model (black).}
\label{fg:sed}
\end{figure}

For an independent, empirical determination of the stellar radius, we analyzed the broadband spectral energy distribution (SED) together with the {\it Gaia\/} parallax, following the procedures described in \cite{stass16,stass17,stass18a}. We obtained the $B_T V_T$ magnitudes from {\it Tycho-2}, the $BVgri$ magnitudes from APASS, the $JHK_S$ magnitudes from {\it 2MASS}, the W1--W4 magnitudes from {\it WISE}, the $G$ magnitude from {\it Gaia\/}, and the FUV and NUV fluxes from {\it GALEX}. Together, the available photometry spans the full stellar SED over the wavelength range 0.15--22~$\mu$m (see Figure~\ref{fg:sed}). 

We performed a fit using Kurucz stellar atmosphere models, with the priors on effective temperature ($T_{\rm eff}$), surface gravity ($\log g$), and metallicity ([Fe/H]) from the spectroscopic parameters listed in Table~\ref{tb:starpar}. The remaining free parameter is the extinction ($A_V$), which we limited to the maximum line-of-sight extinction from the \citet{schle98} dust maps. The resulting fit is very good (Figure~\ref{fg:sed}) with a reduced $\chi^2$ of 3.5, and a best fit extinction of $A_V = 0.08 \pm 0.02$. Integrating the (unextincted) model SED gives the bolometric flux at Earth of $F_{\rm bol} = 3.57 \pm 0.13 \times 10^{-8}$\,erg\,s$^{-1}$\,cm$^{-2}$. Taking the $F_{\rm bol}$ and $T_{\rm eff}$ together with the {\it Gaia\/} parallax, adjusted by $+0.08$\,mas to account for the systematic offset reported by \citet{stasstorres18}, gives the stellar radius as $R_\star = 4.38 \pm 0.20$\,$R_\sun$. 

Combining the bolometric flux with the \textit{Gaia} DR2 distance allows us to derive a nearly model-independent luminosity, which is a valuable constraint for asteroseismic modeling. Using a Gaia parallax ($\pi$) of $11.024 \pm 0.022$\,mas (adjusted for the $0.082 \pm 0.055$ mas zero-point offset for nearby stars reported by \citet{stasstorres18}) with the bolometric flux obtained above yielded $L_\star = 9.14 \pm 0.33\,L_{\sun}$.

In addition, we can estimate the stellar mass ($M_\star$) from the eclipsing-binary based empirical relations of \citet{torres10}, which gives $M_\star = 1.38 \pm 0.14\,\msun$. This, together with the empirical radius above, gives the mean stellar density $\rho_\star = 0.023 \pm 0.004$\,g\,cm$^{-3}$.
However, we note that for $\log g < 3.5$ the empirical relations of \citet{torres10} are extrapolated, therefore the inferred stellar mass for the $\log g$ of HD~222076 should be regarded with caution.

\subsection{\textnormal{TESS} Photometry}
\label{sc:photo}

\begin{figure}
\resizebox{1.0\hsize}{!}{\includegraphics{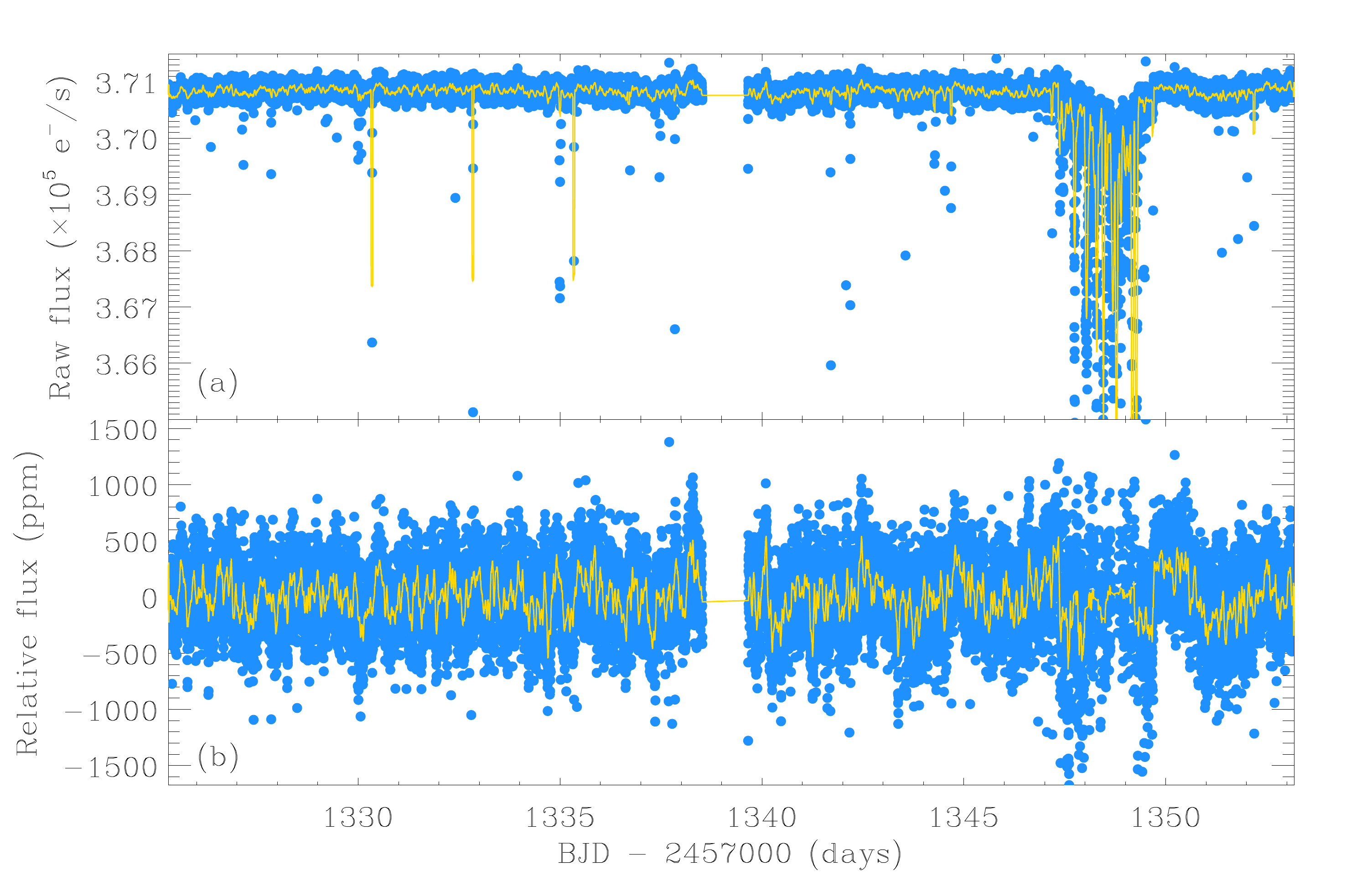}}
\caption{Light curves of HD 222076 produced by the TASOC photometry pipeline. Raw (top) and corrected (bottom) 2-minute cadence light curves are displayed. The yellow lines are the light curve smoothed with a 1-hour boxcar filter (shown for illustration purposes only).}
\label{fg:lightcurve}
\end{figure}

\textit{TESS} observed HD~222076 in 2-minute cadence during Sector 1 of Cycle 1 for a total of 27.9 days. 
According to Figure 5(b) of \cite{camp16b}, with this cadence \textit{TESS} is predicted to detect solar-like oscillations down to an I-band magnitude (used as a proxy for the \textit{TESS} magnitude) of around 10, for a star with effective temperature of $4900\,\rm K$ and luminosity of $9 L_\sun$. Given its \textit{TESS} magnitude. of 6.59, solar-like oscillations are expected to be detected in the light curve of HD~222076.
The light curve was produced using a special version of the photometry pipeline\footnote{https://github.com/tasoc/}\citep{hand19} maintained by the \textit{TESS} Asteroseismic Science Operations Center\footnote{https://tasoc.dk/} \citep[TASOC;][]{lund17b}, which is an extended version of the one adopted in the K2P pipeline \citep{lund15} originally developed to generate light curves for data collected by the K2 Mission. Figure~\ref{fg:lightcurve}(a) shows the raw light curve obtained from the TASOC pipeline. The coverage is nearly continuous with a high duty cycle $\sim$96\%, displaying a $\sim$2 day gap that separates the two spacecraft orbits in the observing sector. A 2.5 days period of high jitter is seen towards the end of the sector, corresponding to instrumental variations due to the spacecraft's angular momentum dumping cycle, which is evident in light curves from Sector 1. The transit of planet HD~222076b is not detected by \textit{TESS} due to the short observation period. For asteroseismic analysis, the raw light curves are subsequently corrected for systematic effects using the KASOC Filter \citep{hand14}. The corrected light curve is shown in Figure~\ref{fg:lightcurve}(b).

\section{Asteroseismic analysis}
\label{sc:seismic}

\subsection{Global Oscillation Parameters}
\label{sc:global}

\begin{figure}
\resizebox{1.0\hsize}{!}{\includegraphics{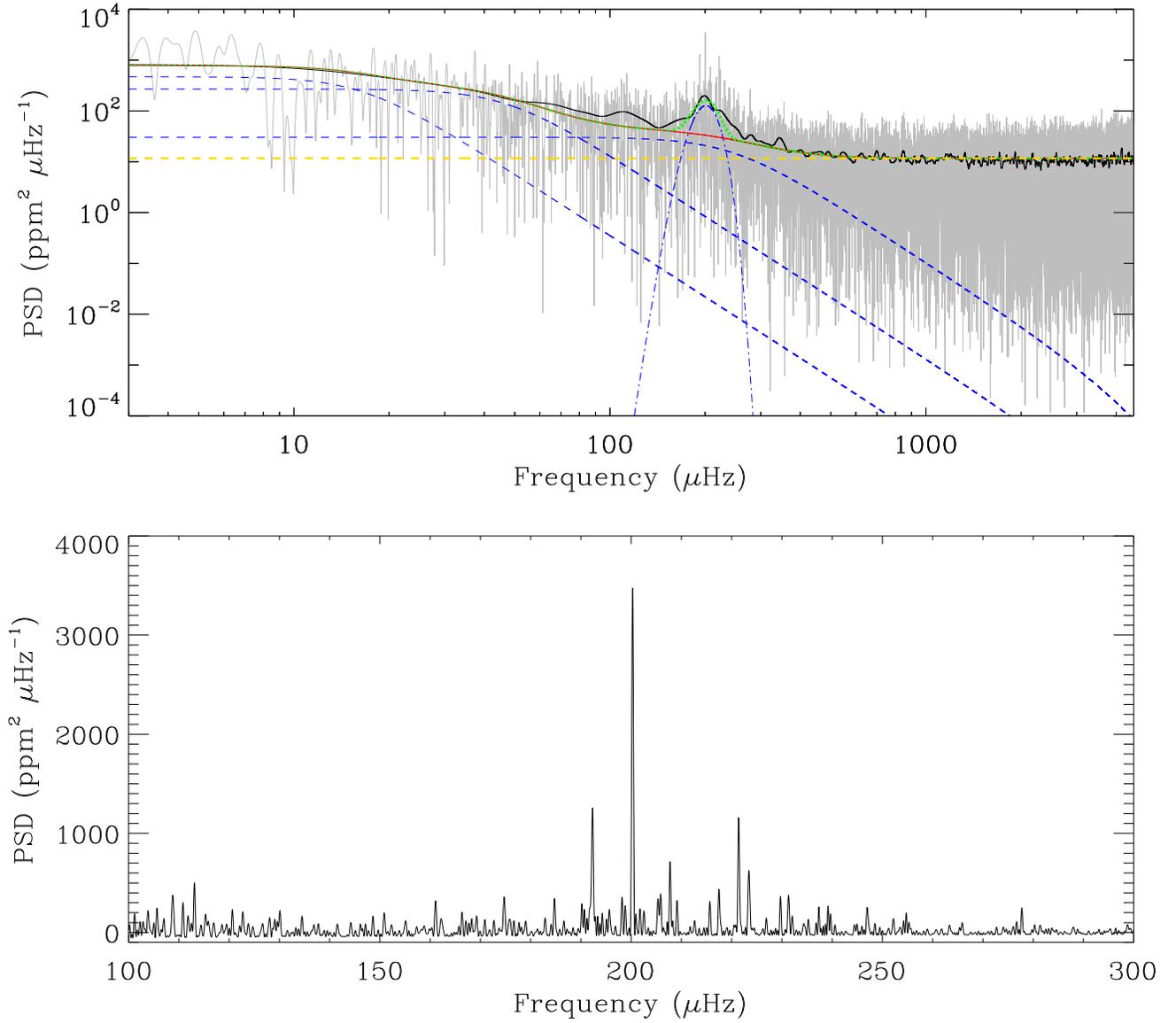}}
\caption{Top panel: power spectral density (PSD) of HD 222076 and corresponding global model fit (green dashed curve). The PSD is shown in gray and a heavily smoothed (Gaussian with an FWHM of $\Dnu$) version in black. The solid red curve is a fit to the background, consisting of three Harvey-like profiles (blue dashed curves) plus the white noise (yellow dashed line). The Gaussian fit to the oscillation power excess is shown by the blue dot-dashed curve. Bottom panel: background-corrected PSD in the range of the stellar oscillations.}
\label{fg:ps}
\end{figure}

To extract oscillation parameters the corrected light curve was analyzed with several different methods \citep{kurtz85, jiang11, corsaro14, corsaro15, buzasi15}, many of which have been extensively applied on \kepler  /K2 data \citep[e.g.,][]{hekker11,verner11}. The top panel of Figure~\ref{fg:ps} illustrates the power spectrum of HD~222076 computed based on the corrected light curve. The power spectrum shows a frequency-dependent background signal due to stellar activity, granulation, and faculae that can be modeled by a sum of several Lorentzian-like functions \citep{harvey85, karoff08, kallinger14, corsaro17}, and a flat noise. The individual components of the background and the final fit using the method by \cite{jiang11} are also shown as dashed blue curves and solid red curve, respectively, in the top panel of Figure~\ref{fg:ps}. Then the total background was subtracted from the power spectrum, and a close-up of the power excess region is shown in the bottom panel of Figure~\ref{fg:ps}.

Next, global seismic parameters such as the frequency of maximum power ($\numax$) and the mean large frequency separation ($\Dnu$) were measured based on the analysis of the background corrected power spectrum. In summary, $\numax$ was measured by fitting a Gaussian function to the the power excess hump of the smoothed power spectrum \citep{huber09, hekker10, mathur10, kallinger14}, as shown in Figure~\ref{fg:ps}. Our analysis yielded $\numax = 203.0 \pm 3.6\, \muHz$. 
To measure $\Dnu$ methods like autocorrelation of the amplitude spectrum \citep{huber09, mosser09}, asymptotic or linear fit to the frequencies of the radial modes (mode degree $\ell = 0$, mode extraction given in Section~\ref{sc:fre}) were used, which gave $\Dnu = 15.60 \pm 0.13\, \muHz$. 
We note that the results from different groups for the two parameters agree within a few percent. A comparison of global oscillation parameters derived from different methods, including the ones used in our analysis, is given by \cite{hekker11}.
The values of $\numax$ and $\Dnu$ were averages over all results reported by different methods. And the uncertainties of the two parameters were calculated from the scatter over all results from different methods. Values for the two parameters are listed in Table~\ref{tb:starpar}.

\subsection{Individual Mode Frequencies}
\label{sc:fre}

The background-corrected power spectrum in Figure~\ref{fg:ps} shows a clear signature of solar-like oscillations: a regular series of peaks spaced by a large separation. Given that \textit{TESS} instrument artifacts are not yet well understood, we restricted our analysis to the frequency range between 150 to 270\,$\muHz$ where we observe peaks well above the noise level. In this region we also see multiple peaks due to dipole mixed modes \citep{beck11, bedding11}.

Individual frequencies were extracted from the power spectrum with several independent methods ranging from traditional iterative fitting of sine waves, i.e., pre-whitening \citep[e.g.][]{hans05, lenz05, bedding07, jiang11}, to fitting of Lorentzian mode profiles \citep[e.g.][]{hand11, app12, mosser12, corsaro14, corsaro15, vrard15, davies16, hand17, rox17, kall18, corsaro19}. Most of the $\ell = 0$ and 2 oscillation modes were successfully identified either based on the frequency ridges in the \'echelle diagram \citep{bedding10} or by using a multi-modal approach presented in \cite{corsaro19}. Very good agreement was achieved from a comparison of the frequencies returned by different methods. 

\begin{figure}
\resizebox{1.0\hsize}{!}{\includegraphics{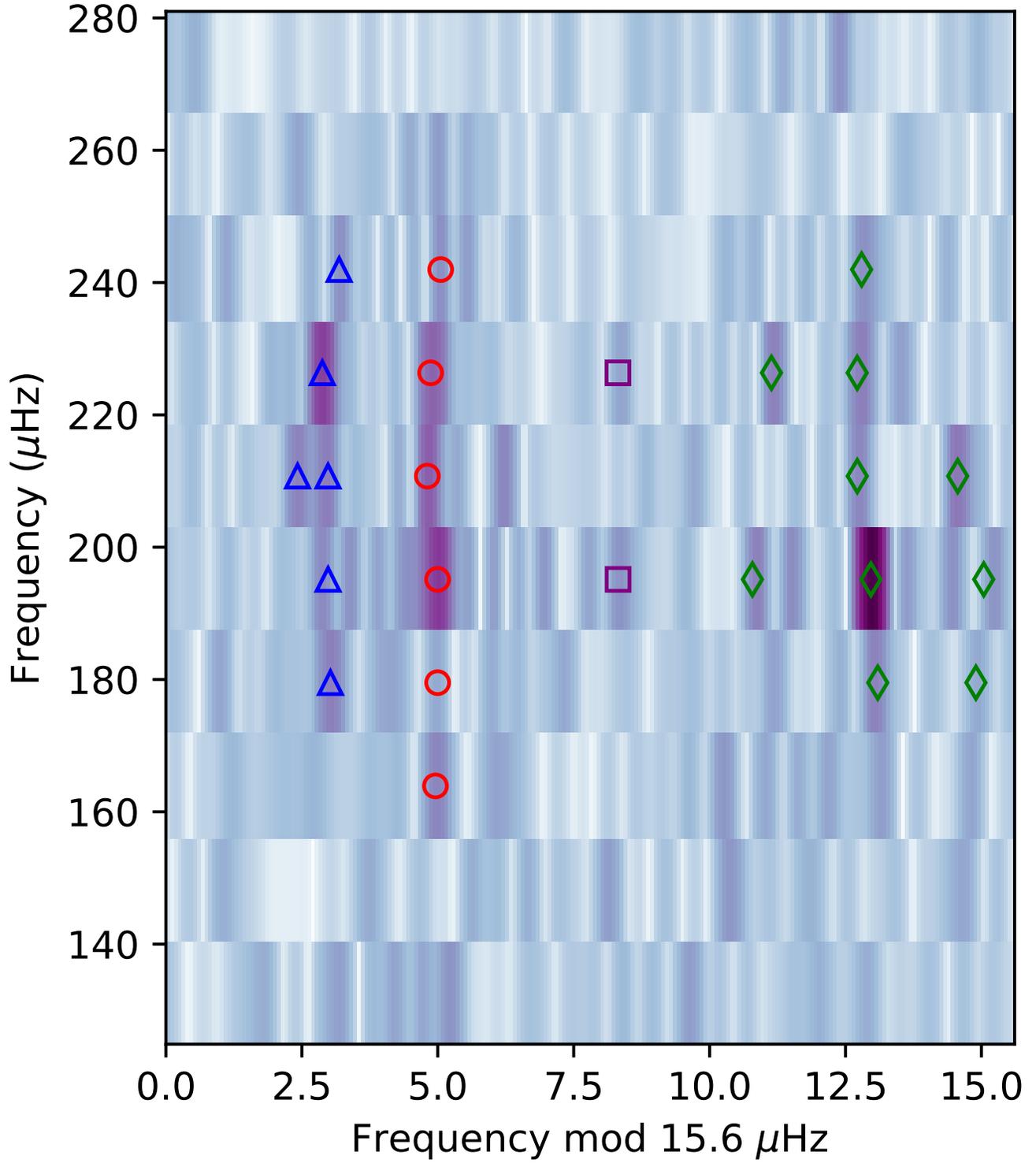}}
\caption{Grayscale \'echelle diagram of the background-corrected PSD. Identified individual mode frequencies are marked with red circles ($\ell =0$), blue triangles ($\ell =2$), green diamonds ($\ell =1$) and purple squares ($\ell =3$). This figure was made using the {\tt echelle} package (Hey \& Ball 2020). }
\label{fg:echelle}
\end{figure}

In Figure~\ref{fg:echelle}, the grayscale \'echelle diagram is illustrated for the background-corrected power spectrum. The identified modes (filled symbols) shown in the figure are returned by at least two independent methods with frequency differences smaller than the uncertainties. However, due to the relatively short observation time of \textit{TESS}, mixed mode patterns are not so clear in the \'echelle diagram. Therefore, the identification of mixed modes needs further confirmation from comparisons with the model frequencies (see Section~\ref{sc:model}). The final frequency list of the identified peaks is given in Table~\ref{tb:frequency}. The radial modes identified from the power spectrum also allowed us to measure $\Dnu$ by performing a linear fit. The resulting value of $\Dnu$ contributes the final estimate given in Section~\ref{sc:global}.

\section{Modeling}
\label{sc:model}

A common way to estimate the fundamental stellar properties is to compare calculated model parameters with the observational constraints that include observed asteroseismic parameters and complementary spectroscopic data. We used a number of independent approaches to model the observed stellar parameters and frequencies, including different stellar evolution codes \citep[ASTEC, MESA;][]{jcd08a, paxton11, paxton13, paxton15}, oscillation codes \citep[ADIPLS, GYRE][]{jcd08b, town13}, and optimization methods \citep[including AIMS, DIAMONDS, PARAM;][]{corsaro14, rod14, wu16, wu17, rod17, frandsen18, nsamba18, zhang18, ong19, randle19, yildiz19}. Corrections for the surface effect \citep{hans08, ball17, viani18} were employed by most of the adopted methods. 
The adopted model inputs included the set of \{[Fe/H], $T_{\rm eff}$, $L_\star$, $\Dnu$, $\nu$\}. The atmospheric parameters ([Fe/H] and $T_{\rm eff}$) are adopted from \cite{witt17}. To investigate the impacts of different inputs, modelers provided results with and without the use of individual frequencies and the luminosity as observable constraints. The diversity of modeling procedures employed implicitly accounts for the impact of using different stellar models and analysis methodologies on the final estimates. While a detailed comparison of the results from different groups is beyond the scope of this paper, we note extensive comparisons of red-giant models and oscillation frequencies computed with 9 widely used stellar evolution codes have recently been performed in the context of \textit{Aarhus Red Giants Challenge} \citep{jcd20, silva20a}.

\begin{figure}
\resizebox{0.8\hsize}{!}{\includegraphics{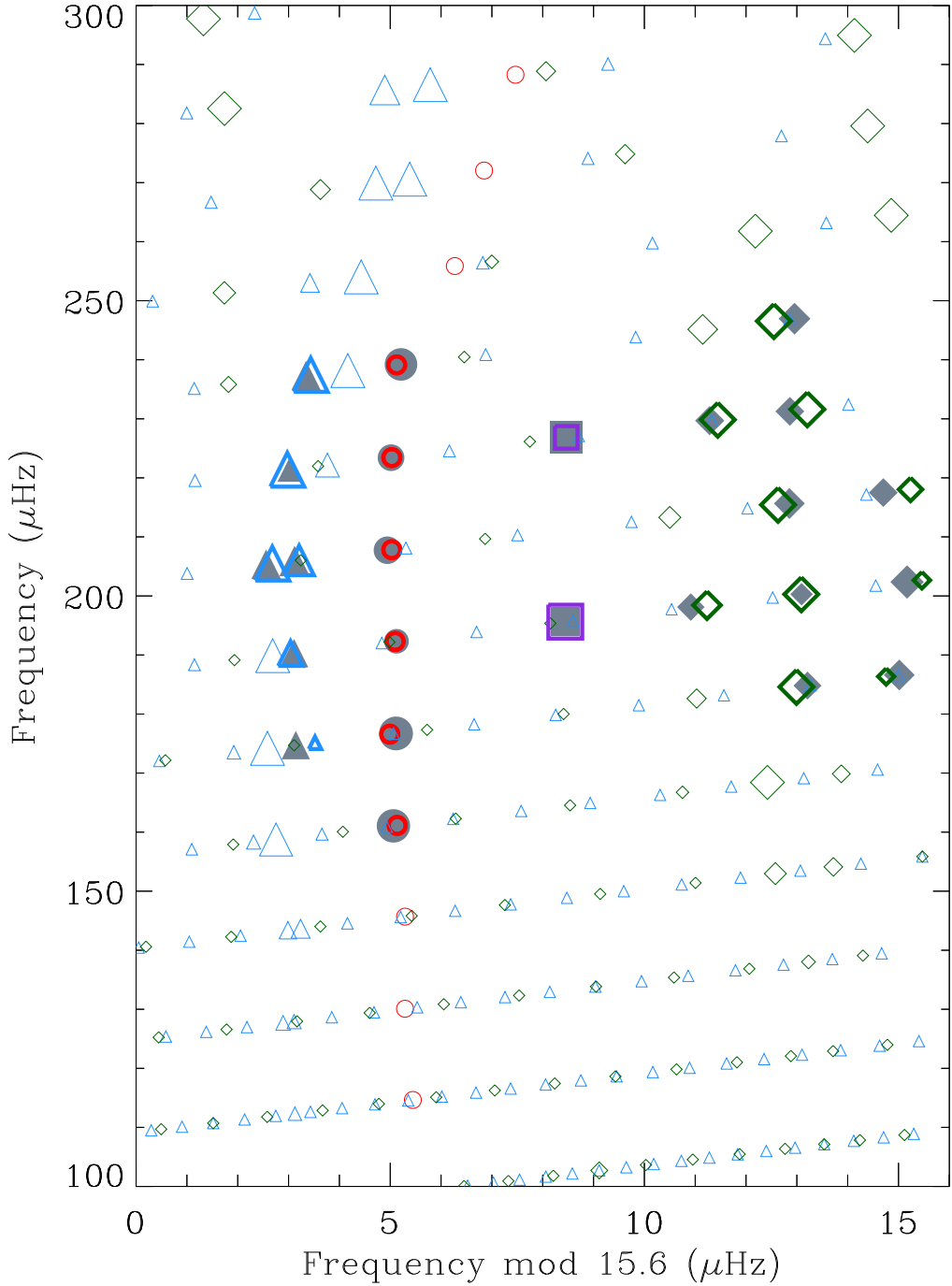}}
\caption{\'Echelle diagram showing observed oscillation frequencies (filled gray symbols) and a representative best-fitting model (open colored symbols) computed by ASTEC and ADIPLS, for $\ell = 0$ (circles), $\ell =1$ (diamonds), $\ell = 2$ (triangles), and $\ell = 3$ (squares) modes. Symbol sizes of observed modes are scaled according to the uncertainties, and those of non-radial theoretical modes are scaled using the inverse inertia as a proxy for mode amplitude \citep{cunha15}. Modes with lower inertial and hence larger symbol sizes have relative higher mode amplitudes. Thick model symbols correspond to modes that are matched to observations. Matched model modes are corrected for the surface effect using the combined correction described by Equation (4) in \cite{ball17}.}
\label{fg:echelle_fit}
\end{figure}
\begin{figure}
\resizebox{1.0\hsize}{!}{\includegraphics{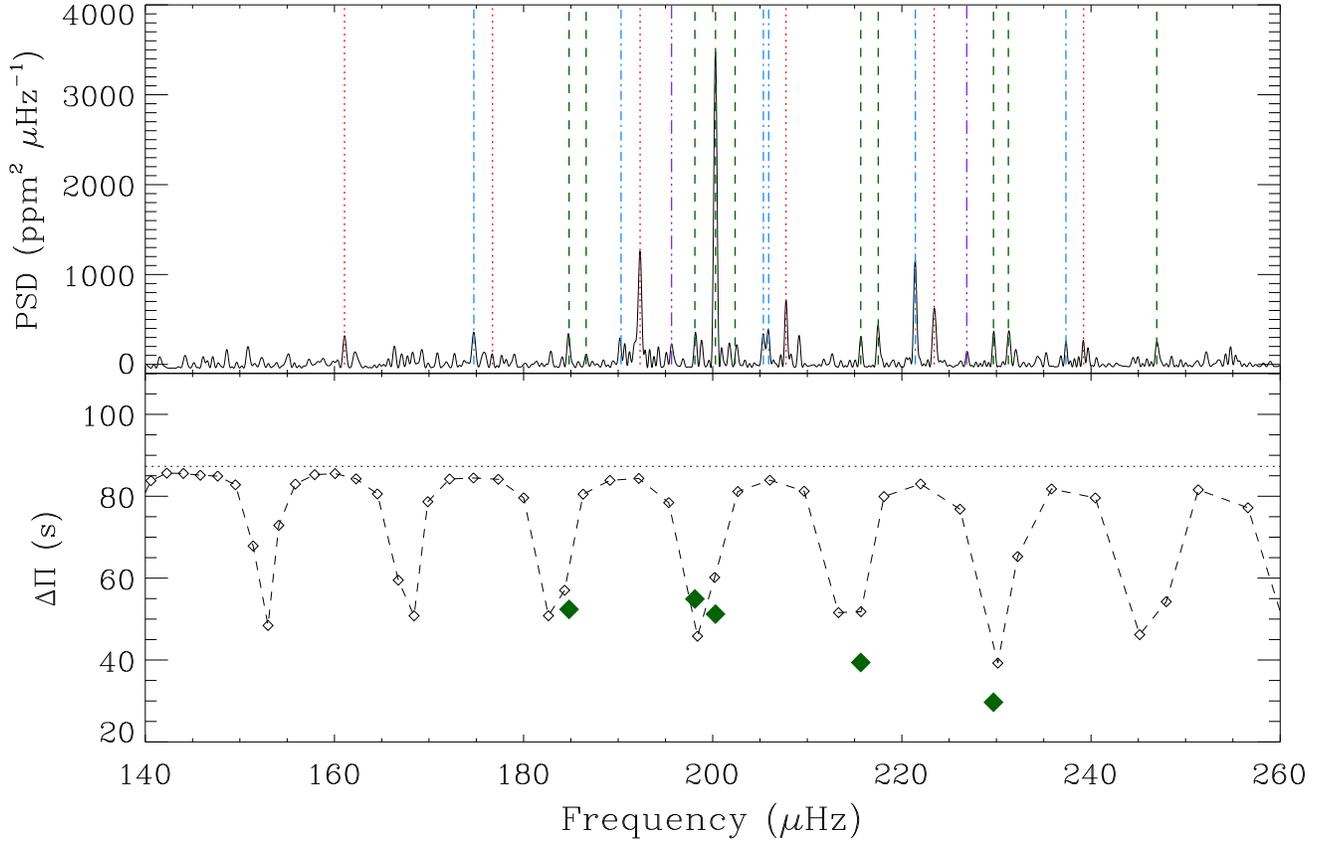}}
\caption{{\em Top}: power density spectrum showing the position of the extracted oscillation peaks identified as $\ell = 0$ (red dotted), $\ell =1$ (green dashed), $\ell = 2$ (blue dash-dotted), and $\ell = 3$ (violet dash-dot-dotted) modes. {\em Bottom}: period spacings $\Dpi$ between adjacent dipolar mixed modes, as a function of the frequency. The dashed line and small open diamonds correspond to the representative best-fitting model shown in Figure~\ref{fg:echelle_fit}. The asymptotic gravity-mode spacing $\Dpi_1$, computed using Equation~\eqref{eq:dpi1}, is indicated by the dotted horizontal line. The filled green diamonds show the period spacings between two observed consecutive modes.}
\label{fg:pspacings}
\end{figure}

Overall, most of the codes were able to find adequate fits to the observable constraints, and the outputs are generally consistent with each other. Most modeling codes were able to provide adequate fits to the observed frequencies. As mentioned before, due to the relative short observational time of \textit{TESS}, the frequency resolution and the peak heights are not good enough, which degrades the possibility to extract the close and the low-amplitude modes from the relative high-noise-level power spectrum. 
The detectability of solar-like oscillations is connected with the ratio of total mean mode power due to acoustic oscillations to the total background power across the frequency range occupied by the oscillations. This quantity provides a global measure of the signal-to-noise ratio (S/N), With \textit{TESS} 2-minute cadence data the total S/N is predicted to be 3.1 for HD~222076, which is obtained based on the formulae in \cite{camp16b} using the noise floor of $60 \, \mathrm{ppm \, hr}^{1/2}$.
By comparing the observed frequencies with those from the best-fitting model (see Figure~\ref{fg:echelle_fit}), we were able to identify the modes with corresponding mode degrees and orders that are also given in Table~\ref{tb:frequency}. 
The oscillation peaks identified from the spectrum (top panel of Figure~\ref{fg:pspacings}) are all with a peak height-to-background ratio larger than 4, exceeding the predicted ratio of 3.1. Our asteroseismic analysis indicates that the \textit{TESS} photometric performance is better than predicted for the case of HD~222076. This is supported by the recently asteroseismic study of 25 \textit{TESS} red giants by \cite{silva20b}. 
Among the extracted modes 12 (10 dipole, 2 quadrupole) are closely spaced mixed modes and 2 are $\ell=3$ modes. 
The identified dipolar mixed modes are pressure-mode-like ones that have larger amplitudes and hence are more likely detectable. 
The asymptotic period spacing of dipolar modes from the best-fitting model, defined by the radial position of the base of the convection zone $r_{\rm bcz}$ as 
\begin{equation}
\Dpi_{1} =  \frac{2\pi^2}{\sqrt{2}} \left(\int_{0}^{r_{\rm bcz}} N_{\rm BV} \frac{\mathrm dr}{r} \right)^{-1},
\label{eq:dpi1}
\end{equation}
is $\Dpi_1 = 87.30$ s. The frequency-independent $\Dpi_1$ is a good model representation of the period spacing \citep{wu19}, usually taken as a constant value, derived from fitting the observed frequencies when the gravity-mode-like mixed modes are dense enough on the power spectrum \citep{mosser14}. In our case with only pressure-mode-like mixed modes obtained from observation, we estimate the observed period spacing $\Dpi_{\rm obs}$, computed as the average spacing of the observed mixed modes (see the bottom panel of Figure~\ref{fg:pspacings}), as $\Dpi_{\rm obs} \approx 45.70 \pm 10.67$ s. The large uncertainty on $\Dpi_{\rm obs}$ is due to the relatively small spacings between the high-frequency modes. However, the values of $\Dpi_{\rm obs}$ and $\Dnu$ indicate that the star HD~222076 is a hydrogen-shell-burning red-giant star \citep{bedding11,corsaro12} that locates before the luminosity bump, and the best-fitting model corroborates that.

The consolidated values for $M_\star$, $R_\star$, $\rho_\star$, $\log g$, and age $t$ of HD~222076 from asteroseismic modeling are summarized in Table~\ref{tb:starpar}, constraining the corresponding properties to $\sim10\%, \sim5\%, \sim6\%, \sim2\%$ and $\sim37\%$, respectively. The uncertainties on these stellar parameters were recalculated by adding the median uncertainty for a given parameter in quadrature to the standard deviation of the parameter estimates returned by all methods. This takes account of both random and systematic errors estimated from different methods and has been commonly adopted for \kepler\ \citep[e.g.,][]{chaplin14b}.  
Adding seismic information in the optimization methods adds extra constraints to the best-fit model selection process. This should yield more precise stellar parameters compared to the empirical eclipsing-binary relation of, e.g., \cite{torres10}. However, the uncertainties on the parameters estimated from the empiric relation presented in Section~\ref{sc:sed} are comparable with those on the asteroseismic values. This is owing to the relatively large uncertainties returned by several modeling methods as well as possible systematic errors reflecting the use of different codes or techniques. Nevertheless, the precision level of stellar parameters obtained in this study resembles that obtained when analyzing the full length of asteroseismic observations from the nominal \kepler\ mission. In addition, the stellar mass from the asteroseismic modeling is expected to be more accurate than that from the empirical relation, which is extrapolated for HD~222076 (cf. Section~\ref{sc:sed}).
The stellar properties estimated in this section indeed have much smaller uncertainties, but are otherwise consistent with those presented in the planet-discovery paper of \cite{witt17}.

\section{Planet Characterization}
\label{sc:planet}

HD~222076b is typical of the population of planets orbiting evolved stars, which are generally beyond $1\,$au and with masses greater than $1\,M_{\rm Jup}$ \citep[e.g.,][]{lovis07,dollinger09,bowler10,jones14,gruetal2019}.  \cite{witt17} computed a lower mass bound of $1.56 \pm 0.11\,M_{\rm Jup}$ and a semimajor axis of $1.83 \pm 0.03$ au for HD 222076~b. We revise this estimate by using the asteroseismic stellar mass ($1.12 \pm 0.12 \,\msun$) in Table 1 and the orbital period ($871 \pm 19\,$days) and velocity semiamplitude ($31.9 \pm 2.3\,$m/s) from \cite{witt17}, and obtain $a = 1.85 \pm 0.07\,$au and $m\sin{i} = 1.62 \pm 0.17\,M_{\rm Jup}$.
We note that by computing the semimajor axis and the minimum mass from the parameters in \cite{witt17} only, we obtain $1.83 \pm 0.14 \,$au and $1.56 \pm 0.28\,M_{\rm Jup}$, which are larger than the ($\pm 0.03\,$au and $\pm 0.11\,M_{\rm Jup}$) uncertainties that they reported. In this sense, our results reduce the uncertainties with the improved stellar mass. 

A semimajor axis of under $2\,$au suggests that a giant planet is in danger of being engulfed during the giant branch phases of stellar evolution \citep{musvil2012,viletal2014,madetal2016,galetal2017,raoetal2018,sunetal2018}. Here, we explore this possibility by applying the same procedure as in \cite{camp19}, which is based on the dynamical tidal formalism of \cite{zahn1977} that was implemented in \cite{viletal2014}, with the added adiabatic assumption of stellar mass loss \citep{veretal2011} and wind velocity and density prescriptions from \cite{veretal2015}. We do not consider atmospheric evaporation \citep{schetal2019}, or the possibility of another hidden planet in the system or any other dynamical process which may affect the planet's post-main-sequence evolution \citep{veras2016}.

We find that the planet is engulfed during the red-giant branch phase and fails to reach the asymptotic giant branch phase. The time at which the engulfment occurs is at about $t = 8.2\,$Gyr, or about $800\,$Myr from now; before the tip of the red-giant branch is reached. This result is insensitive to the choice of planet radius, of which the tidal calculation is a function through the frictional force on the planet (we applied two extreme cases of $1.0\,R_{\rm Jup}$ and $2.0\,R_{\rm Jup}$). The result is, however, sensitive to the planet mass: because the computed value of $1.62\,M_{\rm Jup}$ is a lower limit, this engulfment time represents an upper limit.

\section{Conclusion}

The analysis performed in this work demonstrates the strong potential of \textit{TESS} to characterize exoplanets and their host stars using asteroseismology.  We have re-analyzed the HD~222076 planet system, which was discovered by \cite{witt17}, based on the 2-minute cadence data from \textit{TESS}. The observation time (27.9 days) of the star is rather short compared to the long orbital period (871 days) of the planet, we do not see transit from the data. However, the high-quality photometric observation enables us to perform an asteroseismic analysis of the host star, placing strong constraints on the stellar parameters. From the asteroseismic modeling we obtain a value for the stellar mass of $1.12 \pm 0.12 \, \msun$, a stellar radius of $4.34 \pm 0.21 \, R_{\sun}$ and an age of $7.4 \pm 2.7\,$Gyr. The asteroseismic analysis further allows the detection of 10 dipole mixed modes from the observed power spectrum. The observed period spacing of these mixed modes and the mean large frequency separation reveal that the star HD~222076 is a hydrogen-shell-burning red-giant branch star. Thanks to the measurement of the mixed modes, the evolutionary stage of this star can be analyzed in such level of detail for the first time.

The updated stellar parameters from our asteroseismic analysis have enabled improved estimations for the lower bound of the planetary mass of $m\sin{i} = 1.62 \pm 0.17\,M_{\rm Jup}$ and a semimajor axis of $a = 1.85 \pm 0.07\,$au. With the value obtained for the semimajor axis, we predict that the giant planet is in danger of being engulfed during the giant branch phase of stellar evolution. Based on the estimated stellar age, the engulfment will occur in about $800\,$Myr from now, at the latest.

Our asteroseismic analysis indicates that the \textit{TESS} photometric performance is better than that predicted by \cite{camp16b} for the case of HD~222076. Indeed, \cite{silva20b} found that the quality of \textit{TESS} photometry is similar to that of \kepler\ and K2. This emphasizes the potential of \textit{TESS} for characterizing host stars and understanding their planets.

\section{Acknowledgements}

The project leading to this publication has received funding from the Strategic Priority Research Program of Chinese Academy of Sciences (Grant No. XDB 41000000) and the National Key Program for Science and Technology Research and Development (2017YFB0203300).
This paper includes data collected by the \textit{TESS} mission. Funding for the \textit{TESS} mission is provided by the NASA Explorer Program. Funding for the \textit{TESS} Asteroseismic Science Operations Center at Aarhus University is provided by ESA PRODEX (PEA 4000119301) and Stellar Astrophysics Centre (SAC), funded by the Danish National Research Foundation (Grant agreement No.: DNRF106). 
J.C. is funded by the Fundamental Research Funds for the Central Universities (grant: 19lgpy278) and gratefully thank Biwei Jiang for the discussions on this work.
Q.-S.Z is cosponsored by the National Natural Science Foundation of China (grant No. 11303087) and the Ten Thousands Talents Program of Yunnan Province, and foundations of the Chinese Academy of Sciences (Light of West China Program, Youth Innovation Promotion Association).
D.V. gratefully acknowledges the support of the STFC via an Ernest Rutherford Fellowship (grant ST/P003850/1). 
T.W. and X-Y.Z. thank the supports from the NSFC of China (Grant Nos. 11503076, 11773064, 11873084, and 11521303), from Yunnan Applied Basic Research Projects (Grant No. 2017B008) and from Youth Innovation Promotion Association of Chinese Academy of Sciences. T.W. and X-Y.Z. also gratefully acknowledge the computing time granted by the Yunnan Observatories, and provided on the facilities at the Yunnan Observatories Supercomputing Platform. 
DLB acknowledges support from the Whitaker Center for STEM Education at Florida Gulf Coast University.
B.N. acknowledges postdoctoral funding from the Alexander von Humboldt Foundation taken at the Max-Planck-Institut f\"{u}r Astrophysik (MPA). 
T.D.Li acknowledges support from the Australian Research Council (grant DE~180101104), and the European Research Council (ERC) under the European Union’s Horizon 2020 research and innovation programme (CartographY GA. 804752).
P.M. acknowledges support from NCN grant no. 2016/21/B/ST9/01126.
M.Y., Z.\c{C}.O., and S.\"{O}. acknowedge the Scientific and Technological Research Council of Turkey
(T\"UB\.ITAK:118F352).
TLC acknowledges support from the European Union's Horizon 2020 research and innovation programme under the Marie Sk\l{}odowska-Curie grant agreement No.~792848 (PULSATION). This work was supported by FCT/MCTES through national funds (UID/FIS/04434/2019).
M. S. Cunha is supported by national funds through FCT -- Funda{\c c}\~ao para a Ci\^encia e a Tecnologia - in the form of a work contract and through the research grants UIDB/04434/2020, UIDP/04434/2020 and PTDC/FIS-AST/30389/2017, and by FEDER - Fundo Europeu de Desenvolvimento Regional through COMPETE2020 - Programa Operacional Competitividade e Internacionalização (grant: POCI-01-0145-FEDER-030389).

\bibliography{draft}{}

\begin{thebibliography}{}
\expandafter\ifx\csname natexlab\endcsname\relax\def\natexlab#1{#1}\fi
\providecommand{\url}[1]{\href{#1}{#1}}
\providecommand{\dodoi}[1]{doi:~\href{http://doi.org/#1}{\nolinkurl{#1}}}
\providecommand{\doeprint}[1]{\href{http://ascl.net/#1}{\nolinkurl{http://ascl.net/#1}}}
\providecommand{\doarXiv}[1]{\href{https://arxiv.org/abs/#1}{\nolinkurl{https://arxiv.org/abs/#1}}}

\bibitem[{{Appourchaux} {et~al.}(2012){Appourchaux}, {Chaplin}, {Garc{\'\i}a},
  {Gruberbauer}, {Verner}, {Antia}, {Benomar}, {Campante}, {Davies},
  {Deheuvels}, {Handberg}, {Hekker}, {Howe}, {R{\'e}gulo}, {Salabert},
  {Bedding}, {White}, {Ballot}, {Mathur}, {Silva Aguirre}, {Elsworth}, {Basu},
  {Gilliland }, {Christensen-Dalsgaard}, {Kjeldsen}, {Uddin}, {Stumpe}, \&
  {Barclay}}]{app12}
{Appourchaux}, T., {Chaplin}, W.~J., {Garc{\'\i}a}, R.~A., {et~al.} 2012, \aap,
  543, A54, \dodoi{10.1051/0004-6361/201218948}

\bibitem[{{Baglin} {et~al.}(2006){Baglin}, {Auvergne}, {Boisnard}, {Lam-Trong},
  {Barge}, {Catala}, {Deleuil}, {Michel}, \& {Weiss}}]{baglin06}
{Baglin}, A., {Auvergne}, M., {Boisnard}, L., {et~al.} 2006, in 36th COSPAR
  Scientific Assembly, Vol.~36, 3749

\bibitem[{{Ball} \& {Gizon}(2017)}]{ball17}
{Ball}, W.~H., \& {Gizon}, L. 2017, \aap, 600, A128,
  \dodoi{10.1051/0004-6361/201630260}

\bibitem[{{Ballard} {et~al.}(2014){Ballard}, {Chaplin}, {Charbonneau},
  {D{\'e}sert}, {Fressin}, {Zeng}, {Werner}, {Davies}, {Silva Aguirre}, {Basu},
  {Christensen-Dalsgaard}, {Metcalfe}, {Stello}, {Bedding}, {Campante},
  {Handberg}, {Karoff}, {Elsworth}, {Gilliland}, {Hekker}, {Huber}, {Kawaler},
  {Kjeldsen}, {Lund}, \& {Lundkvist}}]{ballard14}
{Ballard}, S., {Chaplin}, W.~J., {Charbonneau}, D., {et~al.} 2014, \apj, 790,
  12, \dodoi{10.1088/0004-637X/790/1/12}

\bibitem[{{Beck} {et~al.}(2011){Beck}, {Bedding}, {Mosser}, {Stello}, {Garcia},
  {Kallinger}, {Hekker}, {Elsworth}, {Frandsen}, {Carrier}, {De Ridder},
  {Aerts}, {White}, {Huber}, {Dupret}, {Montalb{\'a}n}, {Miglio}, {Noels},
  {Chaplin}, {Kjeldsen}, {Christensen-Dalsgaard}, {Gilliland}, {Brown},
  {Kawaler}, {Mathur}, \& {Jenkins}}]{beck11}
{Beck}, P.~G., {Bedding}, T.~R., {Mosser}, B., {et~al.} 2011, Science, 332,
  205, \dodoi{10.1126/science.1201939}

\bibitem[{{Bedding} \& {Kjeldsen}(2010)}]{bedding10}
{Bedding}, T.~R., \& {Kjeldsen}, H. 2010, Communications in Asteroseismology,
  161, 3, \dodoi{10.1553/cia161s3}

\bibitem[{{Bedding} {et~al.}(2007){Bedding}, {Kjeldsen}, {Arentoft}, {Bouchy},
  {Brandbyge}, {Brewer}, {Butler}, {Christensen-Dalsgaard}, {Dall}, {Frandsen},
  {Karoff}, {Kiss}, {Monteiro}, {Pijpers}, {Teixeira}, {Tinney}, {Baldry},
  {Carrier}, \& {O'Toole}}]{bedding07}
{Bedding}, T.~R., {Kjeldsen}, H., {Arentoft}, T., {et~al.} 2007, \apj, 663,
  1315, \dodoi{10.1086/518593}

\bibitem[{{Bedding} {et~al.}(2011){Bedding}, {Mosser}, {Huber},
  {Montalb{\'a}n}, {Beck}, {Christensen-Dalsgaard}, {Elsworth}, {Garc{\'\i}a},
  {Miglio}, {Stello}, {White}, {De Ridder}, {Hekker}, {Aerts}, {Barban},
  {Belkacem}, {Broomhall}, {Brown}, {Buzasi}, {Carrier}, {Chaplin}, {di Mauro},
  {Dupret}, {Frandsen}, {Gilliland }, {Goupil}, {Jenkins}, {Kallinger},
  {Kawaler}, {Kjeldsen}, {Mathur}, {Noels}, {Silva Aguirre}, \&
  {Ventura}}]{bedding11}
{Bedding}, T.~R., {Mosser}, B., {Huber}, D., {et~al.} 2011, \nat, 471, 608,
  \dodoi{10.1038/nature09935}

\bibitem[{{Benomar} {et~al.}(2014){Benomar}, {Masuda}, {Shibahashi}, \&
  {Suto}}]{benomar14}
{Benomar}, O., {Masuda}, K., {Shibahashi}, H., \& {Suto}, Y. 2014, \pasj, 66,
  94, \dodoi{10.1093/pasj/psu069}

\bibitem[{{Borucki} {et~al.}(2010){Borucki}, {Koch}, {Basri}, {Batalha},
  {Brown}, {Caldwell}, {Caldwell}, {Christensen-Dalsgaard}, {Cochran},
  {DeVore}, {Dunham}, {Dupree}, {Gautier}, {Geary}, {Gilliland}, {Gould},
  {Howell}, {Jenkins}, {Kondo}, {Latham}, {Marcy}, {Meibom}, {Kjeldsen},
  {Lissauer}, {Monet}, {Morrison}, {Sasselov}, {Tarter}, {Boss}, {Brownlee},
  {Owen}, {Buzasi}, {Charbonneau}, {Doyle}, {Fortney}, {Ford}, {Holman},
  {Seager}, {Steffen}, {Welsh}, {Rowe}, {Anderson}, {Buchhave}, {Ciardi},
  {Walkowicz}, {Sherry}, {Horch}, {Isaacson}, {Everett}, {Fischer}, {Torres},
  {Johnson}, {Endl}, {MacQueen}, {Bryson}, {Dotson}, {Haas}, {Kolodziejczak},
  {Van Cleve}, {Chandrasekaran}, {Twicken}, {Quintana}, {Clarke}, {Allen},
  {Li}, {Wu}, {Tenenbaum}, {Verner}, {Bruhweiler}, {Barnes}, \&
  {Prsa}}]{borucki10}
{Borucki}, W.~J., {Koch}, D., {Basri}, G., {et~al.} 2010, Science, 327, 977,
  \dodoi{10.1126/science.1185402}

\bibitem[{{Bowler} {et~al.}(2010){Bowler}, {Johnson}, {Marcy}, {Henry}, {Peek},
  {Fischer}, {Clubb}, {Liu}, {Reffert}, {Schwab}, \& {Lowe}}]{bowler10}
{Bowler}, B.~P., {Johnson}, J.~A., {Marcy}, G.~W., {et~al.} 2010, \apj, 709,
  396, \dodoi{10.1088/0004-637X/709/1/396}

\bibitem[{{Buzasi} {et~al.}(2015){Buzasi}, {Carboneau}, {Hessler}, {Lezcano},
  \& {Preston}}]{buzasi15}
{Buzasi}, D.~L., {Carboneau}, L., {Hessler}, C., {Lezcano}, A., \& {Preston},
  H. 2015, in IAU General Assembly, Vol.~29, 2256843

\bibitem[{{Campante} {et~al.}(2015){Campante}, {Barclay}, {Swift}, {Huber},
  {Adibekyan}, {Cochran}, {Burke}, {Isaacson}, {Quintana}, {Davies}, {Silva
  Aguirre}, {Ragozzine}, {Riddle}, {Baranec}, {Basu}, {Chaplin},
  {Christensen-Dalsgaard}, {Metcalfe}, {Bedding}, {Handberg}, {Stello},
  {Brewer}, {Hekker}, {Karoff}, {Kolbl}, {Law}, {Lundkvist}, {Miglio}, {Rowe},
  {Santos}, {Van Laerhoven}, {Arentoft}, {Elsworth}, {Fischer}, {Kawaler},
  {Kjeldsen}, {Lund}, {Marcy}, {Sousa}, {Sozzetti}, \& {White}}]{camp15}
{Campante}, T.~L., {Barclay}, T., {Swift}, J.~J., {et~al.} 2015, \apj, 799,
  170, \dodoi{10.1088/0004-637X/799/2/170}

\bibitem[{{Campante} {et~al.}(2016a){Campante}, {Lund}, {Kuszlewicz}, {Davies},
  {Chaplin}, {Albrecht}, {Winn}, {Bedding}, {Benomar}, {Bossini}, {Handberg},
  {Santos}, {Van Eylen}, {Basu}, {Christensen-Dalsgaard}, {Elsworth}, {Hekker},
  {Hirano}, {Huber}, {Karoff}, {Kjeldsen}, {Lundkvist}, {North}, {Silva
  Aguirre}, {Stello}, \& {White}}]{camp16a}
{Campante}, T.~L., {Lund}, M.~N., {Kuszlewicz}, J.~S., {et~al.} 2016a, \apj,
  819, 85, \dodoi{10.3847/0004-637X/819/1/85}

\bibitem[{{Campante} {et~al.}(2016b){Campante}, {Schofield}, {Kuszlewicz},
  {Bouma}, {Chaplin}, {Huber}, {Christensen-Dalsgaard}, {Kjeldsen}, {Bossini},
  {North}, {Appourchaux}, {Latham}, {Pepper}, {Ricker}, {Stassun}, {Vand
  erspek}, \& {Winn}}]{camp16b}
{Campante}, T.~L., {Schofield}, M., {Kuszlewicz}, J.~S., {et~al.} 2016b, \apj,
  830, 138, \dodoi{10.3847/0004-637X/830/2/138}

\bibitem[{{Campante} {et~al.}(2019){Campante}, {Corsaro}, {Lund}, {Mosser},
  {Serenelli}, {Veras}, {Adibekyan}, {Antia}, {Ball}, {Basu}, {Bedding},
  {Bossini}, {Davies}, {Delgado Mena}, {Garc{\'\i}a}, {Handberg}, {Hon},
  {Kane}, {Kawaler}, {Kuszlewicz}, {Lucas}, {Mathur}, {Nardetto}, {Nielsen},
  {Pinsonneault}, {Reffert}, {Silva Aguirre}, {Stassun}, {Stello}, {Stock},
  {Vrard}, {Y{\i}ld{\i}z}, {Chaplin}, {Huber}, {Bean}, {{\c{C}}elik Orhan},
  {Cunha}, {Christensen-Dalsgaard}, {Kjeldsen}, {Metcalfe}, {Miglio},
  {Monteiro}, {Nsamba}, {{\"O}rtel}, {Pereira}, {Sousa}, {Tsantaki}, \&
  {Turnbull}}]{camp19}
{Campante}, T.~L., {Corsaro}, E., {Lund}, M.~N., {et~al.} 2019, \apj, 885, 31,
  \dodoi{10.3847/1538-4357/ab44a8}

\bibitem[{{Chaplin} {et~al.}(2014b){Chaplin}, {Elsworth}, {Davies}, {Campante},
  {Handberg}, {Miglio}, \& {Basu}}]{chaplin14b}
{Chaplin}, W.~J., {Elsworth}, Y., {Davies}, G.~R., {et~al.} 2014b, \mnras, 445,
  946, \dodoi{10.1093/mnras/stu1811}

\bibitem[{{Chaplin} \& {Miglio}(2013)}]{chaplin13}
{Chaplin}, W.~J., \& {Miglio}, A. 2013, \araa, 51, 353,
  \dodoi{10.1146/annurev-astro-082812-140938}

\bibitem[{{Chaplin} {et~al.}(2014a){Chaplin}, {Basu}, {Huber}, {Serenelli},
  {Casagrande}, {Silva Aguirre}, {Ball}, {Creevey}, {Gizon}, {Handberg},
  {Karoff}, {Lutz}, {Marques}, {Miglio}, {Stello}, {Suran}, {Pricopi},
  {Metcalfe}, {Monteiro}, {Molenda-{\.Z}akowicz}, {Appourchaux},
  {Christensen-Dalsgaard}, {Elsworth}, {Garc{\'\i}a}, {Houdek}, {Kjeldsen},
  {Bonanno}, {Campante}, {Corsaro}, {Gaulme}, {Hekker}, {Mathur}, {Mosser},
  {R{\'e}gulo}, \& {Salabert}}]{chaplin14a}
{Chaplin}, W.~J., {Basu}, S., {Huber}, D., {et~al.} 2014a, \apjs, 210, 1,
  \dodoi{10.1088/0067-0049/210/1/1}

\bibitem[{{Christensen-Dalsgaard}(2008a)}]{jcd08a}
{Christensen-Dalsgaard}, J. 2008a, \apss, 316, 13,
  \dodoi{10.1007/s10509-007-9675-5}

\bibitem[{{Christensen-Dalsgaard}(2008b)}]{jcd08b}
---. 2008b, \apss, 316, 113, \dodoi{10.1007/s10509-007-9689-z}

\bibitem[{{Christensen-Dalsgaard} {et~al.}(2020){Christensen-Dalsgaard}, {Silva
  Aguirre}, {Cassisi}, {Miller Bertolami}, {Serenelli}, {Stello}, {Weiss},
  {Angelou}, {Jiang}, {Lebreton}, {Spada}, {Bellinger}, {Deheuvels},
  {Ouazzani}, {Pietrinferni}, {Mosumgaard}, {Townsend}, {Battich}, {Bossini},
  {Constantino}, {Eggenberger}, {Hekker}, {Mazumdar}, {Miglio}, {Nielsen}, \&
  {Salaris}}]{jcd20}
{Christensen-Dalsgaard}, J., {Silva Aguirre}, V., {Cassisi}, S., {et~al.} 2020,
  \aap, 635, A165, \dodoi{10.1051/0004-6361/201936766}

\bibitem[{{Corsaro}(2019)}]{corsaro19}
{Corsaro}, E. 2019, Frontiers in Astronomy and Space Sciences, 6, 21,
  \dodoi{10.3389/fspas.2019.00021}

\bibitem[{{Corsaro} \& {De Ridder}(2014)}]{corsaro14}
{Corsaro}, E., \& {De Ridder}, J. 2014, \aap, 571, A71,
  \dodoi{10.1051/0004-6361/201424181}

\bibitem[{{Corsaro} {et~al.}(2015){Corsaro}, {De Ridder}, \&
  {Garc{\'\i}a}}]{corsaro15}
{Corsaro}, E., {De Ridder}, J., \& {Garc{\'\i}a}, R.~A. 2015, \aap, 579, A83,
  \dodoi{10.1051/0004-6361/201525895}

\bibitem[{{Corsaro} {et~al.}(2012){Corsaro}, {Stello}, {Huber}, {Bedding},
  {Bonanno}, {Brogaard}, {Kallinger}, {Benomar}, {White}, {Mosser}, {Basu},
  {Chaplin}, {Christensen-Dalsgaard}, {Elsworth}, {Garc{\'\i}a}, {Hekker},
  {Kjeldsen}, {Mathur}, {Meibom}, {Hall}, {Ibrahim}, \& {Klaus}}]{corsaro12}
{Corsaro}, E., {Stello}, D., {Huber}, D., {et~al.} 2012, \apj, 757, 190,
  \dodoi{10.1088/0004-637X/757/2/190}

\bibitem[{{Corsaro} {et~al.}(2017){Corsaro}, {Mathur}, {Garc{\'\i}a}, {Gaulme},
  {Pinsonneault}, {Stassun}, {Stello}, {Tayar}, {Trampedach}, {Jiang},
  {Nitschelm}, \& {Salabert}}]{corsaro17}
{Corsaro}, E., {Mathur}, S., {Garc{\'\i}a}, R.~A., {et~al.} 2017, \aap, 605,
  A3, \dodoi{10.1051/0004-6361/201731094}

\bibitem[{{Cunha} {et~al.}(2015){Cunha}, {Stello}, {Avelino},
  {Christensen-Dalsgaard}, \& {Townsend}}]{cunha15}
{Cunha}, M.~S., {Stello}, D., {Avelino}, P.~P., {Christensen-Dalsgaard}, J., \&
  {Townsend}, R.~H.~D. 2015, \apj, 805, 127,
  \dodoi{10.1088/0004-637X/805/2/127}

\bibitem[{{Dalba} {et~al.}(2019){Dalba}, {Kane}, {Barclay}, {Bean}, {Campante},
  {Pepper}, {Ragozzine}, \& {Turnbull}}]{dalba19}
{Dalba}, P.~A., {Kane}, S.~R., {Barclay}, T., {et~al.} 2019, \pasp, 131,
  034401, \dodoi{10.1088/1538-3873/aaf183}

\bibitem[{{Davies} \& {Miglio}(2016)}]{davies16}
{Davies}, G.~R., \& {Miglio}, A. 2016, Astronomische Nachrichten, 337, 774,
  \dodoi{10.1002/asna.201612371}

\bibitem[{{Diego} {et~al.}(1990){Diego}, {Charalambous}, {Fish}, \&
  {Walker}}]{diego90}
{Diego}, F., {Charalambous}, A., {Fish}, A.~C., \& {Walker}, D.~D. 1990,
  Society of Photo-Optical Instrumentation Engineers (SPIE) Conference Series,
  Vol. 1235, {Final tests and commissioning of the UCL echelle spectrograph},
  ed. D.~L. {Crawford}, 562--576, \dodoi{10.1117/12.19119}

\bibitem[{{D{\"o}llinger} {et~al.}(2009){D{\"o}llinger}, {Hatzes}, {Pasquini},
  {Guenther}, \& {Hartmann}}]{dollinger09}
{D{\"o}llinger}, M.~P., {Hatzes}, A.~P., {Pasquini}, L., {Guenther}, E.~W., \&
  {Hartmann}, M. 2009, \aap, 505, 1311, \dodoi{10.1051/0004-6361/200911702}

\bibitem[{{Frandsen} {et~al.}(2018){Frandsen}, {Fredslund Andersen},
  {Brogaard}, {Jiang}, {Arentoft}, {Grundahl}, {Kjeldsen},
  {Christensen-Dalsgaard}, {Weiss}, {Pall{\'e}}, {Antoci}, {Kj{\ae}rgaard},
  {S{\o}rensen}, {Skottfelt}, \& {J{\o}rgensen}}]{frandsen18}
{Frandsen}, S., {Fredslund Andersen}, M., {Brogaard}, K., {et~al.} 2018, \aap,
  613, A53, \dodoi{10.1051/0004-6361/201730816}

\bibitem[{{Gaia Collaboration} {et~al.}(2018){Gaia Collaboration}, {Brown},
  {Vallenari}, {Prusti}, {de Bruijne}, {Babusiaux}, {Bailer-Jones}, {Biermann},
  {Evans}, {Eyer}, {Jansen}, {Jordi}, {Klioner}, {Lammers}, {Lindegren},
  {Luri}, {Mignard}, {Panem}, {Pourbaix}, {Randich}, {Sartoretti}, {Siddiqui},
  {Soubiran}, {van Leeuwen}, {Walton}, {Arenou}, {Bastian}, {Cropper},
  {Drimmel}, {Katz}, {Lattanzi}, {Bakker}, {Cacciari}, {Casta{\~n}eda},
  {Chaoul}, {Cheek}, {De Angeli}, {Fabricius}, {Guerra}, {Holl}, {Masana},
  {Messineo}, {Mowlavi}, {Nienartowicz}, {Panuzzo}, {Portell}, {Riello},
  {Seabroke}, {Tanga}, {Th{\'e}venin}, {Gracia-Abril}, {Comoretto},
  {Garcia-Reinaldos}, {Teyssier}, {Altmann}, {Andrae}, {Audard},
  {Bellas-Velidis}, {Benson}, {Berthier}, {Blomme}, {Burgess}, {Busso},
  {Carry}, {Cellino}, {Clementini}, {Clotet}, {Creevey}, {Davidson}, {De
  Ridder}, {Delchambre}, {Dell'Oro}, {Ducourant},
  {Fern{\'a}ndez-Hern{\'a}ndez}, {Fouesneau}, {Fr{\'e}mat}, {Galluccio},
  {Garc{\'\i}a-Torres}, {Gonz{\'a}lez-N{\'u}{\~n}ez}, {Gonz{\'a}lez-Vidal},
  {Gosset}, {Guy}, {Halbwachs}, {Hambly}, {Harrison}, {Hern{\'a}ndez},
  {Hestroffer}, {Hodgkin}, {Hutton}, {Jasniewicz}, {Jean-Antoine-Piccolo},
  {Jordan}, {Korn}, {Krone-Martins}, {Lanzafame}, {Lebzelter}, {L{\"o}ffler},
  {Manteiga}, {Marrese}, {Mart{\'\i}n-Fleitas}, {Moitinho}, {Mora}, {Muinonen},
  {Osinde}, {Pancino}, {Pauwels}, {Petit}, {Recio-Blanco}, {Richards},
  {Rimoldini}, {Robin}, {Sarro}, {Siopis}, {Smith}, {Sozzetti}, {S{\"u}veges},
  {Torra}, {van Reeven}, {Abbas}, {Abreu Aramburu}, {Accart}, {Aerts},
  {Altavilla}, {{\'A}lvarez}, {Alvarez}, {Alves}, {Anderson}, {Andrei},
  {Anglada Varela}, {Antiche}, {Antoja}, {Arcay}, {Astraatmadja}, {Bach},
  {Baker}, {Balaguer-N{\'u}{\~n}ez}, {Balm}, {Barache}, {Barata}, {Barbato},
  {Barblan}, {Barklem}, {Barrado}, {Barros}, {Barstow}, {Bartholom{\'e}
  Mu{\~n}oz}, {Bassilana}, {Becciani}, {Bellazzini}, {Berihuete}, {Bertone},
  {Bianchi}, {Bienaym{\'e}}, {Blanco-Cuaresma}, {Boch}, {Boeche}, {Bombrun},
  {Borrachero}, {Bossini}, {Bouquillon}, {Bourda}, {Bragaglia}, {Bramante},
  {Breddels}, {Bressan}, {Brouillet}, {Br{\"u}semeister}, {Brugaletta},
  {Bucciarelli}, {Burlacu}, {Busonero}, {Butkevich}, {Buzzi}, {Caffau},
  {Cancelliere}, {Cannizzaro}, {Cantat-Gaudin}, {Carballo}, {Carlucci},
  {Carrasco}, {Casamiquela}, {Castellani}, {Castro-Ginard}, {Charlot},
  {Chemin}, {Chiavassa}, {Cocozza}, {Costigan}, {Cowell}, {Crifo}, {Crosta},
  {Crowley}, {Cuypers}, {Dafonte}, {Damerdji}, {Dapergolas}, {David}, {David},
  {de Laverny}, {De Luise}, {De March}, {de Martino}, {de Souza}, {de Torres},
  {Debosscher}, {del Pozo}, {Delbo}, {Delgado}, {Delgado}, {Di Matteo},
  {Diakite}, {Diener}, {Distefano}, {Dolding}, {Drazinos}, {Dur{\'a}n},
  {Edvardsson}, {Enke}, {Eriksson}, {Esquej}, {Eynard Bontemps}, {Fabre},
  {Fabrizio}, {Faigler}, {Falc{\~a}o}, {Farr{\`a}s Casas}, {Federici},
  {Fedorets}, {Fernique}, {Figueras}, {Filippi}, {Findeisen}, {Fonti},
  {Fraile}, {Fraser}, {Fr{\'e}zouls}, {Gai}, {Galleti}, {Garabato},
  {Garc{\'\i}a-Sedano}, {Garofalo}, {Garralda}, {Gavel}, {Gavras}, {Gerssen},
  {Geyer}, {Giacobbe}, {Gilmore}, {Girona}, {Giuffrida}, {Glass}, {Gomes},
  {Granvik}, {Gueguen}, {Guerrier}, {Guiraud}, {Guti{\'e}rrez-S{\'a}nchez},
  {Haigron}, {Hatzidimitriou}, {Hauser}, {Haywood}, {Heiter}, {Helmi}, {Heu},
  {Hilger}, {Hobbs}, {Hofmann}, {Holland}, {Huckle}, {Hypki}, {Icardi},
  {Jan{\ss}en}, {Jevardat de Fombelle}, {Jonker}, {Juh{\'a}sz}, {Julbe},
  {Karampelas}, {Kewley}, {Klar}, {Kochoska}, {Kohley}, {Kolenberg},
  {Kontizas}, {Kontizas}, {Koposov}, {Kordopatis}, {Kostrzewa-Rutkowska},
  {Koubsky}, {Lambert}, {Lanza}, {Lasne}, {Lavigne}, {Le Fustec}, {Le
  Poncin-Lafitte}, {Lebreton}, {Leccia}, {Leclerc}, {Lecoeur-Taibi},
  {Lenhardt}, {Leroux}, {Liao}, {Licata}, {Lindstr{\o}m}, {Lister}, {Livanou},
  {Lobel}, {L{\'o}pez}, {Managau}, {Mann}, {Mantelet}, {Marchal}, {Marchant},
  {Marconi}, {Marinoni}, {Marschalk{\'o}}, {Marshall}, {Martino}, {Marton},
  {Mary}, {Massari}, {Matijevi{\v{c}}}, {Mazeh}, {McMillan}, {Messina},
  {Michalik}, {Millar}, {Molina}, {Molinaro}, {Moln{\'a}r}, {Montegriffo},
  {Mor}, {Morbidelli}, {Morel}, {Morris}, {Mulone}, {Muraveva}, {Musella},
  {Nelemans}, {Nicastro}, {Noval}, {O'Mullane}, {Ord{\'e}novic},
  {Ord{\'o}{\~n}ez-Blanco}, {Osborne}, {Pagani}, {Pagano}, {Pailler},
  {Palacin}, {Palaversa}, {Panahi}, {Pawlak}, {Piersimoni}, {Pineau}, {Plachy},
  {Plum}, {Poggio}, {Poujoulet}, {Pr{\v{s}}a}, {Pulone}, {Racero}, {Ragaini},
  {Rambaux}, {Ramos-Lerate}, {Regibo}, {Reyl{\'e}}, {Riclet}, {Ripepi}, {Riva},
  {Rivard}, {Rixon}, {Roegiers}, {Roelens}, {Romero-G{\'o}mez}, {Rowell},
  {Royer}, {Ruiz-Dern}, {Sadowski}, {Sagrist{\`a} Sell{\'e}s}, {Sahlmann},
  {Salgado}, {Salguero}, {Sanna}, {Santana-Ros}, {Sarasso}, {Savietto},
  {Schultheis}, {Sciacca}, {Segol}, {Segovia}, {S{\'e}gransan}, {Shih},
  {Siltala}, {Silva}, {Smart}, {Smith}, {Solano}, {Solitro}, {Sordo}, {Soria
  Nieto}, {Souchay}, {Spagna}, {Spoto}, {Stampa}, {Steele},
  {Steidelm{\"u}ller}, {Stephenson}, {Stoev}, {Suess}, {Surdej}, {Szabados},
  {Szegedi-Elek}, {Tapiador}, {Taris}, {Tauran}, {Taylor}, {Teixeira},
  {Terrett}, {Teyssand ier}, {Thuillot}, {Titarenko}, {Torra Clotet}, {Turon},
  {Ulla}, {Utrilla}, {Uzzi}, {Vaillant}, {Valentini}, {Valette}, {van Elteren},
  {Van Hemelryck}, {van Leeuwen}, {Vaschetto}, {Vecchiato}, {Veljanoski},
  {Viala}, {Vicente}, {Vogt}, {von Essen}, {Voss}, {Votruba}, {Voutsinas},
  {Walmsley}, {Weiler}, {Wertz}, {Wevers}, {Wyrzykowski}, {Yoldas},
  {{\v{Z}}erjal}, {Ziaeepour}, {Zorec}, {Zschocke}, {Zucker}, {Zurbach}, \&
  {Zwitter}}]{GaiaDR2}
{Gaia Collaboration}, {Brown}, A.~G.~A., {Vallenari}, A., {et~al.} 2018, \aap,
  616, A1, \dodoi{10.1051/0004-6361/201833051}

\bibitem[{{Gallet} {et~al.}(2017){Gallet}, {Bolmont}, {Mathis}, {Charbonnel},
  \& {Amard}}]{galetal2017}
{Gallet}, F., {Bolmont}, E., {Mathis}, S., {Charbonnel}, C., \& {Amard}, L.
  2017, \aap, 604, A112, \dodoi{10.1051/0004-6361/201730661}

\bibitem[{{Grunblatt} {et~al.}(2019){Grunblatt}, {Huber}, {Gaidos}, {Hon},
  {Zinn}, \& {Stello}}]{gruetal2019}
{Grunblatt}, S.~K., {Huber}, D., {Gaidos}, E., {et~al.} 2019, \aj, 158, 227,
  \dodoi{10.3847/1538-3881/ab4c35}

\bibitem[{{Handberg} {et~al.}(2017){Handberg}, {Brogaard}, {Miglio}, {Bossini},
  {Elsworth}, {Slumstrup}, {Davies}, \& {Chaplin}}]{hand17}
{Handberg}, R., {Brogaard}, K., {Miglio}, A., {et~al.} 2017, \mnras, 472, 979,
  \dodoi{10.1093/mnras/stx1929}

\bibitem[{{Handberg} \& {Campante}(2011)}]{hand11}
{Handberg}, R., \& {Campante}, T.~L. 2011, \aap, 527, A56,
  \dodoi{10.1051/0004-6361/201015451}

\bibitem[{{Handberg} \& {Lund}(2014)}]{hand14}
{Handberg}, R., \& {Lund}, M.~N. 2014, \mnras, 445, 2698,
  \dodoi{10.1093/mnras/stu1823}

\bibitem[{Handberg \& Lund(2019)}]{hand19}
Handberg, R., \& Lund, M.~N. 2019, {T'DA Data Release Notes - Data Release 4
  for TESS Sectors 1 + 2}, \dodoi{10.5281/zenodo.2579846}

\bibitem[{{Harvey}(1985)}]{harvey85}
{Harvey}, J. 1985, in ESA Special Publication, Vol. 235, Future Missions in
  Solar, Heliospheric \& Space Plasma Physics, ed. E.~{Rolfe} \& B.~{Battrick},
  199

\bibitem[{{Hekker} {et~al.}(2010){Hekker}, {Broomhall}, {Chaplin}, {Elsworth},
  {Fletcher}, {New}, {Arentoft}, {Quirion}, \& {Kjeldsen}}]{hekker10}
{Hekker}, S., {Broomhall}, A.~M., {Chaplin}, W.~J., {et~al.} 2010, \mnras, 402,
  2049, \dodoi{10.1111/j.1365-2966.2009.16030.x}

\bibitem[{{Hekker} {et~al.}(2011){Hekker}, {Elsworth}, {De Ridder}, {Mosser},
  {Garc{\'\i}a}, {Kallinger}, {Mathur}, {Huber}, {Buzasi}, {Preston}, {Hale},
  {Ballot}, {Chaplin}, {R{\'e}gulo}, {Bedding}, {Stello}, {Borucki}, {Koch},
  {Jenkins}, {Allen}, {Gilliland}, {Kjeldsen}, \&
  {Christensen-Dalsgaard}}]{hekker11}
{Hekker}, S., {Elsworth}, Y., {De Ridder}, J., {et~al.} 2011, \aap, 525, A131,
  \dodoi{10.1051/0004-6361/201015185}

\bibitem[{{Houk} \& {Cowley}(1975)}]{houk75}
{Houk}, N., \& {Cowley}, A.~P. 1975, {University of Michigan Catalogue of
  two-dimensional spectral types for the HD stars. Volume I. Declinations
  -90{\fdg}0 to -53{\fdg}0.}

\bibitem[{{Huber} {et~al.}(2009){Huber}, {Stello}, {Bedding}, {Chaplin},
  {Arentoft}, {Quirion}, \& {Kjeldsen}}]{huber09}
{Huber}, D., {Stello}, D., {Bedding}, T.~R., {et~al.} 2009, Communications in
  Asteroseismology, 160, 74.
\newblock \doarXiv{0910.2764}

\bibitem[{{Huber} {et~al.}(2013){Huber}, {Carter}, {Barbieri}, {Miglio},
  {Deck}, {Fabrycky}, {Montet}, {Buchhave}, {Chaplin}, {Hekker},
  {Montalb{\'a}n}, {Sanchis-Ojeda}, {Basu}, {Bedding}, {Campante},
  {Christensen-Dalsgaard}, {Elsworth}, {Stello}, {Arentoft}, {Ford}, {Gilliland
  }, {Handberg}, {Howard}, {Isaacson}, {Johnson}, {Karoff}, {Kawaler},
  {Kjeldsen}, {Latham}, {Lund}, {Lundkvist}, {Marcy}, {Metcalfe}, {Silva
  Aguirre}, \& {Winn}}]{huber13}
{Huber}, D., {Carter}, J.~A., {Barbieri}, M., {et~al.} 2013, Science, 342, 331,
  \dodoi{10.1126/science.1242066}

\bibitem[{{Huber} {et~al.}(2019){Huber}, {Chaplin}, {Chontos}, {Kjeldsen},
  {Christensen-Dalsgaard}, {Bedding}, {Ball}, {Brahm}, {Espinoza}, {Henning},
  {Jord{\'a}n}, {Sarkis}, {Knudstrup}, {Albrecht}, {Grundahl}, {Fredslund
  Andersen}, {Pall{\'e}}, {Crossfield}, {Fulton}, {Howard}, {Isaacson},
  {Weiss}, {Handberg}, {Lund}, {Serenelli}, {R{\o}rsted Mosumgaard},
  {Stokholm}, {Bieryla}, {Buchhave}, {Latham}, {Quinn}, {Gaidos}, {Hirano},
  {Ricker}, {Vanderspek}, {Seager}, {Jenkins}, {Winn}, {Antia}, {Appourchaux},
  {Basu}, {Bell}, {Benomar}, {Bonanno}, {Buzasi}, {Campante}, {{\c{C}}elik
  Orhan}, {Corsaro}, {Cunha}, {Davies}, {Deheuvels}, {Grunblatt}, {Hasanzadeh},
  {Di Mauro}, {Garc{\'\i}a}, {Gaulme}, {Girardi}, {Guzik}, {Hon}, {Jiang},
  {Kallinger}, {Kawaler}, {Kuszlewicz}, {Lebreton}, {Li}, {Lucas}, {Lundkvist},
  {Mann}, {Mathis}, {Mathur}, {Mazumdar}, {Metcalfe}, {Miglio}, {Monteiro},
  {Mosser}, {Noll}, {Nsamba}, {Ong}, {{\"O}rtel}, {Pereira}, {Ranadive},
  {R{\'e}gulo}, {Rodrigues}, {Roxburgh}, {Silva Aguirre}, {Smalley},
  {Schofield}, {Sousa}, {Stassun}, {Stello}, {Tayar}, {White}, {Verma},
  {Vrard}, {Y{\i}ld{\i}z}, {Baker}, {Bazot}, {Beichmann}, {Bergmann}, {Bugnet},
  {Cale}, {Carlino}, {Cartwright}, {Christiansen}, {Ciardi}, {Creevey},
  {Dittmann}, {Do Nascimento}, {Van Eylen}, {F{\"u}r{\'e}sz}, {Gagn{\'e}},
  {Gao}, {Gazeas}, {Giddens}, {Hall}, {Hekker}, {Ireland }, {Latouf}, {LeBrun},
  {Levine}, {Matzko}, {Natinsky}, {Page}, {Plavchan}, {Mansouri-Samani},
  {McCauliff}, {Mullally}, {Orenstein}, {Garcia Soto}, {Paegert}, {van Saders},
  {Schnaible}, {Soderblom}, {Szab{\'o}}, {Tanner}, {Tinney}, {Teske}, {Thomas},
  {Trampedach}, {Wright}, {Yuan}, \& {Zohrabi}}]{huber19}
{Huber}, D., {Chaplin}, W.~J., {Chontos}, A., {et~al.} 2019, \aj, 157, 245,
  \dodoi{10.3847/1538-3881/ab1488}

\bibitem[{{Jiang} {et~al.}(2011){Jiang}, {Jiang}, {Christensen-Dalsgaard},
  {Bedding}, {Stello}, {Huber}, {Frandsen}, {Kjeldsen}, {Karoff}, {Mosser},
  {Demarque}, {Fanelli}, {Kinemuchi}, \& {Mullally}}]{jiang11}
{Jiang}, C., {Jiang}, B.~W., {Christensen-Dalsgaard}, J., {et~al.} 2011, \apj,
  742, 120, \dodoi{10.1088/0004-637X/742/2/120}

\bibitem[{{Jones} {et~al.}(2014){Jones}, {Jenkins}, {Bluhm}, {Rojo}, \&
  {Melo}}]{jones14}
{Jones}, M.~I., {Jenkins}, J.~S., {Bluhm}, P., {Rojo}, P., \& {Melo}, C.~H.~F.
  2014, \aap, 566, A113, \dodoi{10.1051/0004-6361/201323345}

\bibitem[{{Jones} {et~al.}(2011){Jones}, {Jenkins}, {Rojo}, \&
  {Melo}}]{jones11}
{Jones}, M.~I., {Jenkins}, J.~S., {Rojo}, P., \& {Melo}, C.~H.~F. 2011, \aap,
  536, A71, \dodoi{10.1051/0004-6361/201117887}

\bibitem[{{Kallinger} {et~al.}(2018){Kallinger}, {Beck}, {Stello}, \&
  {Garcia}}]{kall18}
{Kallinger}, T., {Beck}, P.~G., {Stello}, D., \& {Garcia}, R.~A. 2018, \aap,
  616, A104, \dodoi{10.1051/0004-6361/201832831}

\bibitem[{{Kallinger} {et~al.}(2014){Kallinger}, {De Ridder}, {Hekker},
  {Mathur}, {Mosser}, {Gruberbauer}, {Garc{\'\i}a}, {Karoff}, \&
  {Ballot}}]{kallinger14}
{Kallinger}, T., {De Ridder}, J., {Hekker}, S., {et~al.} 2014, \aap, 570, A41,
  \dodoi{10.1051/0004-6361/201424313}

\bibitem[{{Kamiaka} {et~al.}(2019){Kamiaka}, {Benomar}, {Suto}, {Dai},
  {Masuda}, \& {Winn}}]{kamiaka19}
{Kamiaka}, S., {Benomar}, O., {Suto}, Y., {et~al.} 2019, \aj, 157, 137,
  \dodoi{10.3847/1538-3881/ab04a9}

\bibitem[{{Kane} {et~al.}(2012){Kane}, {Ciardi}, {Gelino}, \& {von
  Braun}}]{kane12}
{Kane}, S.~R., {Ciardi}, D.~R., {Gelino}, D.~M., \& {von Braun}, K. 2012,
  \mnras, 425, 757, \dodoi{10.1111/j.1365-2966.2012.21627.x}

\bibitem[{{Karoff}(2008)}]{karoff08}
{Karoff}, C. 2008, PhD thesis, Aarhus University

\bibitem[{{Kaufer} {et~al.}(1999){Kaufer}, {Stahl}, {Tubbesing},
  {N{\o}rregaard}, {Avila}, {Francois}, {Pasquini}, \& {Pizzella}}]{kaufer99}
{Kaufer}, A., {Stahl}, O., {Tubbesing}, S., {et~al.} 1999, The Messenger, 95, 8

\bibitem[{{Kjeldsen} {et~al.}(2008){Kjeldsen}, {Bedding}, \&
  {Christensen-Dalsgaard}}]{hans08}
{Kjeldsen}, H., {Bedding}, T.~R., \& {Christensen-Dalsgaard}, J. 2008, \apjl,
  683, L175, \dodoi{10.1086/591667}

\bibitem[{{Kjeldsen} {et~al.}(2005){Kjeldsen}, {Bedding}, {Butler},
  {Christensen-Dalsgaard}, {Kiss}, {McCarthy}, {Marcy}, {Tinney}, \&
  {Wright}}]{hans05}
{Kjeldsen}, H., {Bedding}, T.~R., {Butler}, R.~P., {et~al.} 2005, \apj, 635,
  1281, \dodoi{10.1086/497530}

\bibitem[{{Kurtz}(1985)}]{kurtz85}
{Kurtz}, D.~W. 1985, \mnras, 213, 773, \dodoi{10.1093/mnras/213.4.773}

\bibitem[{{Lenz} \& {Breger}(2005)}]{lenz05}
{Lenz}, P., \& {Breger}, M. 2005, Communications in Asteroseismology, 146, 53,
  \dodoi{10.1553/cia146s53}

\bibitem[{{Lovis} \& {Mayor}(2007)}]{lovis07}
{Lovis}, C., \& {Mayor}, M. 2007, \aap, 472, 657,
  \dodoi{10.1051/0004-6361:20077375}

\bibitem[{{Lund} {et~al.}(2015){Lund}, {Handberg}, {Davies}, {Chaplin}, \&
  {Jones}}]{lund15}
{Lund}, M.~N., {Handberg}, R., {Davies}, G.~R., {Chaplin}, W.~J., \& {Jones},
  C.~D. 2015, \apj, 806, 30, \dodoi{10.1088/0004-637X/806/1/30}

\bibitem[{{Lund} {et~al.}(2017b){Lund}, {Handberg}, {Kjeldsen}, {Chaplin}, \&
  {Christensen-Dalsgaard}}]{lund17b}
{Lund}, M.~N., {Handberg}, R., {Kjeldsen}, H., {Chaplin}, W.~J., \&
  {Christensen-Dalsgaard}, J. 2017b, in European Physical Journal Web of
  Conferences, Vol. 160, European Physical Journal Web of Conferences, 01005,
  \dodoi{10.1051/epjconf/201716001005}

\bibitem[{{Lund} {et~al.}(2014){Lund}, {Lundkvist}, {Silva Aguirre}, {Houdek},
  {Casagrande}, {Van Eylen}, {Campante}, {Karoff}, {Kjeldsen}, {Albrecht},
  {Chaplin}, {Nielsen}, {Degroote}, {Davies}, \& {Handberg}}]{lund14}
{Lund}, M.~N., {Lundkvist}, M., {Silva Aguirre}, V., {et~al.} 2014, \aap, 570,
  A54, \dodoi{10.1051/0004-6361/201424326}

\bibitem[{{Lund} {et~al.}(2017a){Lund}, {Silva Aguirre}, {Davies}, {Chaplin},
  {Christensen-Dalsgaard}, {Houdek}, {White}, {Bedding}, {Ball}, {Huber},
  {Antia}, {Lebreton}, {Latham}, {Handberg}, {Verma}, {Basu}, {Casagrande},
  {Justesen}, {Kjeldsen}, \& {Mosumgaard}}]{lund17a}
{Lund}, M.~N., {Silva Aguirre}, V., {Davies}, G.~R., {et~al.} 2017a, \apj, 835,
  172, \dodoi{10.3847/1538-4357/835/2/172}

\bibitem[{{Lundkvist} {et~al.}(2016){Lundkvist}, {Kjeldsen}, {Albrecht},
  {Davies}, {Basu}, {Huber}, {Justesen}, {Karoff}, {Silva Aguirre}, {van
  Eylen}, {Vang}, {Arentoft}, {Barclay}, {Bedding}, {Campante}, {Chaplin},
  {Christensen-Dalsgaard}, {Elsworth}, {Gilliland}, {Handberg}, {Hekker},
  {Kawaler}, {Lund}, {Metcalfe}, {Miglio}, {Rowe}, {Stello}, {Tingley}, \&
  {White}}]{lundkvist16}
{Lundkvist}, M.~S., {Kjeldsen}, H., {Albrecht}, S., {et~al.} 2016, Nature
  Communications, 7, 11201, \dodoi{10.1038/ncomms11201}

\bibitem[{{Madappatt} {et~al.}(2016){Madappatt}, {De Marco}, \&
  {Villaver}}]{madetal2016}
{Madappatt}, N., {De Marco}, O., \& {Villaver}, E. 2016, \mnras, 463, 1040,
  \dodoi{10.1093/mnras/stw2025}

\bibitem[{{Mathur} {et~al.}(2010){Mathur}, {Garc{\'\i}a}, {R{\'e}gulo},
  {Creevey}, {Ballot}, {Salabert}, {Arentoft}, {Quirion}, {Chaplin}, \&
  {Kjeldsen}}]{mathur10}
{Mathur}, S., {Garc{\'\i}a}, R.~A., {R{\'e}gulo}, C., {et~al.} 2010, \aap, 511,
  A46, \dodoi{10.1051/0004-6361/200913266}

\bibitem[{{Mathur} {et~al.}(2012){Mathur}, {Metcalfe}, {Woitaszek}, {Bruntt},
  {Verner}, {Christensen-Dalsgaard}, {Creevey}, {Do{\v{g}}an}, {Basu},
  {Karoff}, {Stello}, {Appourchaux}, {Campante}, {Chaplin}, {Garc{\'\i}a},
  {Bedding}, {Benomar}, {Bonanno}, {Deheuvels}, {Elsworth}, {Gaulme}, {Guzik},
  {Handberg}, {Hekker}, {Herzberg}, {Monteiro}, {Piau}, {Quirion},
  {R{\'e}gulo}, {Roth}, {Salabert}, {Serenelli}, {Thompson}, {Trampedach},
  {White}, {Ballot}, {Brand{\~a}o}, {Molenda-{\.Z}akowicz}, {Kjeldsen},
  {Twicken}, {Uddin}, \& {Wohler}}]{mathur12}
{Mathur}, S., {Metcalfe}, T.~S., {Woitaszek}, M., {et~al.} 2012, \apj, 749,
  152, \dodoi{10.1088/0004-637X/749/2/152}

\bibitem[{{Metcalfe} {et~al.}(2010){Metcalfe}, {Monteiro}, {Thompson},
  {Molenda-{\.Z}akowicz}, {Appourchaux}, {Chaplin}, {Do{\v{g}}an},
  {Eggenberger}, {Bedding}, {Bruntt}, {Creevey}, {Quirion}, {Stello},
  {Bonanno}, {Silva Aguirre}, {Basu}, {Esch}, {Gai}, {Di Mauro}, {Kosovichev},
  {Kitiashvili}, {Su{\'a}rez}, {Moya}, {Piau}, {Garc{\'\i}a}, {Marques},
  {Frasca}, {Biazzo}, {Sousa}, {Dreizler}, {Bazot}, {Karoff}, {Frandsen},
  {Wilson}, {Brown}, {Christensen-Dalsgaard}, {Gilliland}, {Kjeldsen},
  {Campante}, {Fletcher}, {Hand berg}, {R{\'e}gulo}, {Salabert}, {Schou},
  {Verner}, {Ballot}, {Broomhall}, {Elsworth}, {Hekker}, {Huber}, {Mathur},
  {New}, {Roxburgh}, {Sato}, {White}, {Borucki}, {Koch}, \&
  {Jenkins}}]{metcalfe10}
{Metcalfe}, T.~S., {Monteiro}, M.~J.~P.~F.~G., {Thompson}, M.~J., {et~al.}
  2010, \apj, 723, 1583, \dodoi{10.1088/0004-637X/723/2/1583}

\bibitem[{{Mosser} \& {Appourchaux}(2009)}]{mosser09}
{Mosser}, B., \& {Appourchaux}, T. 2009, \aap, 508, 877,
  \dodoi{10.1051/0004-6361/200912944}

\bibitem[{{Mosser} {et~al.}(2012){Mosser}, {Goupil}, {Belkacem}, {Michel},
  {Stello}, {Marques}, {Elsworth}, {Barban}, {Beck}, {Bedding}, {De Ridder},
  {Garc{\'\i}a}, {Hekker}, {Kallinger}, {Samadi}, {Stumpe}, {Barclay}, \&
  {Burke}}]{mosser12}
{Mosser}, B., {Goupil}, M.~J., {Belkacem}, K., {et~al.} 2012, \aap, 540, A143,
  \dodoi{10.1051/0004-6361/201118519}

\bibitem[{{Mosser} {et~al.}(2014){Mosser}, {Benomar}, {Belkacem}, {Goupil},
  {Lagarde}, {Michel}, {Lebreton}, {Stello}, {Vrard}, {Barban}, {Bedding},
  {Deheuvels}, {Chaplin}, {De Ridder}, {Elsworth}, {Montalban}, {Noels},
  {Ouazzani}, {Samadi}, {White}, \& {Kjeldsen}}]{mosser14}
{Mosser}, B., {Benomar}, O., {Belkacem}, K., {et~al.} 2014, \aap, 572, L5,
  \dodoi{10.1051/0004-6361/201425039}

\bibitem[{{Mustill} \& {Villaver}(2012)}]{musvil2012}
{Mustill}, A.~J., \& {Villaver}, E. 2012, \apj, 761, 121,
  \dodoi{10.1088/0004-637X/761/2/121}

\bibitem[{{Nsamba} {et~al.}(2018){Nsamba}, {Campante}, {Monteiro}, {Cunha},
  {Rendle}, {Reese}, \& {Verma}}]{nsamba18}
{Nsamba}, B., {Campante}, T.~L., {Monteiro}, M.~J.~P.~F.~G., {et~al.} 2018,
  \mnras, 477, 5052, \dodoi{10.1093/mnras/sty948}

\bibitem[{{Ong} \& {Basu}(2019)}]{ong19}
{Ong}, J.~M.~J., \& {Basu}, S. 2019, \apj, 870, 41,
  \dodoi{10.3847/1538-4357/aaf1b5}

\bibitem[{{Paxton} {et~al.}(2011){Paxton}, {Bildsten}, {Dotter}, {Herwig},
  {Lesaffre}, \& {Timmes}}]{paxton11}
{Paxton}, B., {Bildsten}, L., {Dotter}, A., {et~al.} 2011, \apjs, 192, 3,
  \dodoi{10.1088/0067-0049/192/1/3}

\bibitem[{{Paxton} {et~al.}(2013){Paxton}, {Cantiello}, {Arras}, {Bildsten},
  {Brown}, {Dotter}, {Mankovich}, {Montgomery}, {Stello}, {Timmes}, \&
  {Townsend}}]{paxton13}
{Paxton}, B., {Cantiello}, M., {Arras}, P., {et~al.} 2013, \apjs, 208, 4,
  \dodoi{10.1088/0067-0049/208/1/4}

\bibitem[{{Paxton} {et~al.}(2015){Paxton}, {Marchant}, {Schwab}, {Bauer},
  {Bildsten}, {Cantiello}, {Dessart}, {Farmer}, {Hu}, {Langer}, {Townsend},
  {Townsley}, \& {Timmes}}]{paxton15}
{Paxton}, B., {Marchant}, P., {Schwab}, J., {et~al.} 2015, \apjs, 220, 15,
  \dodoi{10.1088/0067-0049/220/1/15}

\bibitem[{{Rao} {et~al.}(2018){Rao}, {Meynet}, {Eggenberger}, {Haemmerl{\'e}},
  {Privitera}, {Georgy}, {Ekstr{\"o}m}, \& {Mordasini}}]{raoetal2018}
{Rao}, S., {Meynet}, G., {Eggenberger}, P., {et~al.} 2018, \aap, 618, A18,
  \dodoi{10.1051/0004-6361/201833107}

\bibitem[{{Rendle} {et~al.}(2019){Rendle}, {Buldgen}, {Miglio}, {Reese},
  {Noels}, {Davies}, {Campante}, {Chaplin}, {Lund}, {Kuszlewicz}, {Scott},
  {Scuflaire}, {Ball}, {Smetana}, \& {Nsamba}}]{randle19}
{Rendle}, B.~M., {Buldgen}, G., {Miglio}, A., {et~al.} 2019, \mnras, 484, 771,
  \dodoi{10.1093/mnras/stz031}

\bibitem[{{Ricker} {et~al.}(2014){Ricker}, {Winn}, {Vanderspek}, {Latham},
  {Bakos}, {Bean}, {Berta-Thompson}, {Brown}, {Buchhave}, {Butler}, {Butler},
  {Chaplin}, {Charbonneau}, {Christensen-Dalsgaard}, {Clampin}, {Deming},
  {Doty}, {De Lee}, {Dressing}, {Dunham}, {Endl}, {Fressin}, {Ge}, {Henning},
  {Holman}, {Howard}, {Ida}, {Jenkins}, {Jernigan}, {Johnson}, {Kaltenegger},
  {Kawai}, {Kjeldsen}, {Laughlin}, {Levine}, {Lin}, {Lissauer}, {MacQueen},
  {Marcy}, {McCullough}, {Morton}, {Narita}, {Paegert}, {Palle}, {Pepe},
  {Pepper}, {Quirrenbach}, {Rinehart}, {Sasselov}, {Sato}, {Seager},
  {Sozzetti}, {Stassun}, {Sullivan}, {Szentgyorgyi}, {Torres}, {Udry}, \&
  {Villasenor}}]{ricker14}
{Ricker}, G.~R., {Winn}, J.~N., {Vanderspek}, R., {et~al.} 2014, Society of
  Photo-Optical Instrumentation Engineers (SPIE) Conference Series, Vol. 9143,
  {Transiting Exoplanet Survey Satellite (TESS)}, 914320,
  \dodoi{10.1117/12.2063489}

\bibitem[{{Rodrigues} {et~al.}(2014){Rodrigues}, {Girardi}, {Miglio},
  {Bossini}, {Bovy}, {Epstein}, {Pinsonneault}, {Stello}, {Zasowski}, {Allende
  Prieto}, {Chaplin}, {Hekker}, {Johnson}, {M{\'e}sz{\'a}ros}, {Mosser},
  {Anders}, {Basu}, {Beers}, {Chiappini}, {da Costa}, {Elsworth},
  {Garc{\'\i}a}, {Garc{\'\i}a P{\'e}rez}, {Hearty}, {Maia}, {Majewski},
  {Mathur}, {Montalb{\'a}n}, {Nidever}, {Santiago}, {Schultheis}, {Serenelli},
  \& {Shetrone}}]{rod14}
{Rodrigues}, T.~S., {Girardi}, L., {Miglio}, A., {et~al.} 2014, \mnras, 445,
  2758, \dodoi{10.1093/mnras/stu1907}

\bibitem[{{Rodrigues} {et~al.}(2017){Rodrigues}, {Bossini}, {Miglio},
  {Girardi}, {Montalb{\'a}n}, {Noels}, {Trabucchi}, {Coelho}, \&
  {Marigo}}]{rod17}
{Rodrigues}, T.~S., {Bossini}, D., {Miglio}, A., {et~al.} 2017, \mnras, 467,
  1433, \dodoi{10.1093/mnras/stx120}

\bibitem[{{Roxburgh}(2017)}]{rox17}
{Roxburgh}, I.~W. 2017, \aap, 604, A42, \dodoi{10.1051/0004-6361/201731057}

\bibitem[{{Santos} {et~al.}(2013){Santos}, {Sousa}, {Mortier}, {Neves},
  {Adibekyan}, {Tsantaki}, {Delgado Mena}, {Bonfils}, {Israelian}, {Mayor}, \&
  {Udry}}]{santos13}
{Santos}, N.~C., {Sousa}, S.~G., {Mortier}, A., {et~al.} 2013, \aap, 556, A150,
  \dodoi{10.1051/0004-6361/201321286}

\bibitem[{{Schlegel} {et~al.}(1998){Schlegel}, {Finkbeiner}, \&
  {Davis}}]{schle98}
{Schlegel}, D.~J., {Finkbeiner}, D.~P., \& {Davis}, M. 1998, \apj, 500, 525,
  \dodoi{10.1086/305772}

\bibitem[{{Schofield} {et~al.}(2019){Schofield}, {Chaplin}, {Huber},
  {Campante}, {Davies}, {Miglio}, {Ball}, {Appourchaux}, {Basu}, {Bedding},
  {Christensen-Dalsgaard}, {Creevey}, {Garc{\'\i}a}, {Handberg}, {Kawaler},
  {Kjeldsen}, {Latham}, {Lund}, {Metcalfe}, {Ricker}, {Serenelli}, {Silva
  Aguirre}, {Stello}, \& {Vanderspek}}]{schofield19}
{Schofield}, M., {Chaplin}, W.~J., {Huber}, D., {et~al.} 2019, \apjs, 241, 12,
  \dodoi{10.3847/1538-4365/ab04f5}

\bibitem[{{Schreiber} {et~al.}(2019){Schreiber}, {G{\"a}nsicke}, {Toloza},
  {Hernandez}, \& {Lagos}}]{schetal2019}
{Schreiber}, M.~R., {G{\"a}nsicke}, B.~T., {Toloza}, O., {Hernandez}, M.-S., \&
  {Lagos}, F. 2019, \apjl, 887, L4, \dodoi{10.3847/2041-8213/ab42e2}

\bibitem[{{Serenelli} {et~al.}(2017){Serenelli}, {Johnson}, {Huber},
  {Pinsonneault}, {Ball}, {Tayar}, {Silva Aguirre}, {Basu}, {Troup}, {Hekker},
  {Kallinger}, {Stello}, {Davies}, {Lund}, {Mathur}, {Mosser}, {Stassun},
  {Chaplin}, {Elsworth}, {Garc{\'\i}a}, {Handberg}, {Holtzman}, {Hearty},
  {Garc{\'\i}a-Hern{\'a}ndez}, {Gaulme}, \& {Zamora}}]{seren17}
{Serenelli}, A., {Johnson}, J., {Huber}, D., {et~al.} 2017, \apjs, 233, 23,
  \dodoi{10.3847/1538-4365/aa97df}

\bibitem[{{Silva Aguirre} {et~al.}(2015){Silva Aguirre}, {Davies}, {Basu},
  {Christensen-Dalsgaard}, {Creevey}, {Metcalfe}, {Bedding}, {Casagrande},
  {Handberg}, {Lund}, {Nissen}, {Chaplin}, {Huber}, {Serenelli}, {Stello}, {Van
  Eylen}, {Campante}, {Elsworth}, {Gilliland}, {Hekker}, {Karoff}, {Kawaler},
  {Kjeldsen}, \& {Lundkvist}}]{silva15}
{Silva Aguirre}, V., {Davies}, G.~R., {Basu}, S., {et~al.} 2015, \mnras, 452,
  2127, \dodoi{10.1093/mnras/stv1388}

\bibitem[{{Silva Aguirre} {et~al.}(2017){Silva Aguirre}, {Lund}, {Antia},
  {Ball}, {Basu}, {Christensen-Dalsgaard}, {Lebreton}, {Reese}, {Verma},
  {Casagrande}, {Justesen}, {Mosumgaard}, {Chaplin}, {Bedding}, {Davies},
  {Handberg}, {Houdek}, {Huber}, {Kjeldsen}, {Latham}, {White}, {Coelho},
  {Miglio}, \& {Rendle}}]{silva17}
{Silva Aguirre}, V., {Lund}, M.~N., {Antia}, H.~M., {et~al.} 2017, \apj, 835,
  173, \dodoi{10.3847/1538-4357/835/2/173}

\bibitem[{{Silva Aguirre} {et~al.}(2020a){Silva Aguirre},
  {Christensen-Dalsgaard}, {Cassisi}, {Miller Bertolami}, {Serenelli},
  {Stello}, {Weiss}, {Angelou}, {Jiang}, {Lebreton}, {Spada}, {Bellinger},
  {Deheuvels}, {Ouazzani}, {Pietrinferni}, {Mosumgaard}, {Townsend}, {Battich},
  {Bossini}, {Constantino}, {Eggenberger}, {Hekker}, {Mazumdar}, {Miglio},
  {Nielsen}, \& {Salaris}}]{silva20a}
{Silva Aguirre}, V., {Christensen-Dalsgaard}, J., {Cassisi}, S., {et~al.}
  2020a, \aap, 635, A164, \dodoi{10.1051/0004-6361/201935843}

\bibitem[{{Silva Aguirre} {et~al.}(2020b){Silva Aguirre}, {Stello}, {Stokholm},
  {Mosumgaard}, {Ball}, {Basu}, {Bossini}, {Bugnet}, {Buzasi}, {Campante},
  {Carboneau}, {Chaplin}, {Corsaro}, {Davies}, {Elsworth}, {Garc{\'\i}a},
  {Gaulme}, {Hall}, {Handberg}, {Hon}, {Kallinger}, {Kang}, {Lund}, {Mathur},
  {Mints}, {Mosser}, {{\c{C}}elik Orhan}, {Rodrigues}, {Vrard}, {Y{\i}ld{\i}z},
  {Zinn}, {{\"O}rtel}, {Beck}, {Bell}, {Guo}, {Jiang}, {Kuszlewicz}, {Kuehn},
  {Li}, {Lundkvist}, {Pinsonneault}, {Tayar}, {Cunha}, {Hekker}, {Huber},
  {Miglio}, {F.~G. Monteiro}, {Slumstrup}, {Winther}, {Angelou}, {Benomar},
  {B{\'o}di}, {De Moura}, {Deheuvels}, {Derekas}, {Di Mauro}, {Dupret},
  {Jim{\'e}nez}, {Lebreton}, {Matthews}, {Nardetto}, {do Nascimento},
  {Pereira}, {Rodr{\'\i}guez D{\'\i}az}, {Serenelli}, {Spitoni},
  {Stonkut{\.{e}}}, {Su{\'a}rez}, {Szab{\'o}}, {Van Eylen}, {Ventura}, {Verma},
  {Weiss}, {Wu}, {Barclay}, {Christensen-Dalsgaard}, {Jenkins}, {Kjeldsen},
  {Ricker}, {Seager}, \& {Vanderspek}}]{silva20b}
{Silva Aguirre}, V., {Stello}, D., {Stokholm}, A., {et~al.} 2020b, \apjl, 889,
  L34, \dodoi{10.3847/2041-8213/ab6443}

\bibitem[{{Sliski} \& {Kipping}(2014)}]{sliski14}
{Sliski}, D.~H., \& {Kipping}, D.~M. 2014, \apj, 788, 148,
  \dodoi{10.1088/0004-637X/788/2/148}

\bibitem[{{Sousa} {et~al.}(2018){Sousa}, {Adibekyan}, {Delgado-Mena}, {Santos},
  {Andreasen}, {Ferreira}, {Tsantaki}, {Barros}, {Demangeon}, {Israelian},
  {Faria}, {Figueira}, {Mortier}, {Brand{\~a}o}, {Montalto}, {Rojas-Ayala}, \&
  {Santerne}}]{sousa18}
{Sousa}, S.~G., {Adibekyan}, V., {Delgado-Mena}, E., {et~al.} 2018, \aap, 620,
  A58, \dodoi{10.1051/0004-6361/201833350}

\bibitem[{{Stassun} {et~al.}(2017){Stassun}, {Collins}, \& {Gaudi}}]{stass17}
{Stassun}, K.~G., {Collins}, K.~A., \& {Gaudi}, B.~S. 2017, \aj, 153, 136,
  \dodoi{10.3847/1538-3881/aa5df3}

\bibitem[{{Stassun} {et~al.}(2018a){Stassun}, {Corsaro}, {Pepper}, \&
  {Gaudi}}]{stass18a}
{Stassun}, K.~G., {Corsaro}, E., {Pepper}, J.~A., \& {Gaudi}, B.~S. 2018a, \aj,
  155, 22, \dodoi{10.3847/1538-3881/aa998a}

\bibitem[{{Stassun} \& {Torres}(2016)}]{stass16}
{Stassun}, K.~G., \& {Torres}, G. 2016, \aj, 152, 180,
  \dodoi{10.3847/0004-6256/152/6/180}

\bibitem[{{Stassun} \& {Torres}(2018)}]{stasstorres18}
---. 2018, \apj, 862, 61, \dodoi{10.3847/1538-4357/aacafc}

\bibitem[{{Stassun} {et~al.}(2018b){Stassun}, {Oelkers}, {Pepper}, {Paegert},
  {De Lee}, {Torres}, {Latham}, {Charpinet}, {Dressing}, {Huber}, {Kane},
  {L{\'e}pine}, {Mann}, {Muirhead}, {Rojas-Ayala}, {Silvotti}, {Fleming},
  {Levine}, \& {Plavchan}}]{stass18b}
{Stassun}, K.~G., {Oelkers}, R.~J., {Pepper}, J., {et~al.} 2018b, \aj, 156,
  102, \dodoi{10.3847/1538-3881/aad050}

\bibitem[{{Sun} {et~al.}(2018){Sun}, {Arras}, {Weinberg}, {Troup}, \&
  {Majewski}}]{sunetal2018}
{Sun}, M., {Arras}, P., {Weinberg}, N.~N., {Troup}, N.~W., \& {Majewski}, S.~R.
  2018, \mnras, 481, 4077, \dodoi{10.1093/mnras/sty2464}

\bibitem[{{Tokovinin} {et~al.}(2013){Tokovinin}, {Fischer}, {Bonati},
  {Giguere}, {Moore}, {Schwab}, {Spronck}, \& {Szymkowiak}}]{toko13}
{Tokovinin}, A., {Fischer}, D.~A., {Bonati}, M., {et~al.} 2013, \pasp, 125,
  1336, \dodoi{10.1086/674012}

\bibitem[{{Torres} {et~al.}(2010){Torres}, {Andersen}, \&
  {Gim{\'e}nez}}]{torres10}
{Torres}, G., {Andersen}, J., \& {Gim{\'e}nez}, A. 2010, \aapr, 18, 67,
  \dodoi{10.1007/s00159-009-0025-1}

\bibitem[{{Townsend} \& {Teitler}(2013)}]{town13}
{Townsend}, R.~H.~D., \& {Teitler}, S.~A. 2013, \mnras, 435, 3406,
  \dodoi{10.1093/mnras/stt1533}

\bibitem[{{Van Eylen} \& {Albrecht}(2015)}]{van15}
{Van Eylen}, V., \& {Albrecht}, S. 2015, \apj, 808, 126,
  \dodoi{10.1088/0004-637X/808/2/126}

\bibitem[{{Van Eylen} {et~al.}(2019){Van Eylen}, {Albrecht}, {Huang},
  {MacDonald}, {Dawson}, {Cai}, {Foreman-Mackey}, {Lundkvist}, {Silva Aguirre},
  {Snellen}, \& {Winn}}]{van19}
{Van Eylen}, V., {Albrecht}, S., {Huang}, X., {et~al.} 2019, \aj, 157, 61,
  \dodoi{10.3847/1538-3881/aaf22f}

\bibitem[{{van Leeuwen}(2007)}]{HIP}
{van Leeuwen}, F. 2007, \aap, 474, 653, \dodoi{10.1051/0004-6361:20078357}

\bibitem[{{Veras}(2016)}]{veras2016}
{Veras}, D. 2016, Royal Society Open Science, 3, 150571,
  \dodoi{10.1098/rsos.150571}

\bibitem[{{Veras} {et~al.}(2015){Veras}, {Eggl}, \&
  {G{\"a}nsicke}}]{veretal2015}
{Veras}, D., {Eggl}, S., \& {G{\"a}nsicke}, B.~T. 2015, \mnras, 451, 2814,
  \dodoi{10.1093/mnras/stv1047}

\bibitem[{{Veras} {et~al.}(2011){Veras}, {Wyatt}, {Mustill}, {Bonsor}, \&
  {Eldridge}}]{veretal2011}
{Veras}, D., {Wyatt}, M.~C., {Mustill}, A.~J., {Bonsor}, A., \& {Eldridge},
  J.~J. 2011, \mnras, 417, 2104, \dodoi{10.1111/j.1365-2966.2011.19393.x}

\bibitem[{{Verner} {et~al.}(2011){Verner}, {Elsworth}, {Chaplin}, {Campante},
  {Corsaro}, {Gaulme}, {Hekker}, {Huber}, {Karoff}, {Mathur}, {Mosser},
  {Appourchaux}, {Ballot}, {Bedding}, {Bonanno}, {Broomhall}, {Garc{\'\i}a},
  {Hand berg}, {New}, {Stello}, {R{\'e}gulo}, {Roxburgh}, {Salabert}, {White},
  {Caldwell}, {Christiansen}, \& {Fanelli}}]{verner11}
{Verner}, G.~A., {Elsworth}, Y., {Chaplin}, W.~J., {et~al.} 2011, \mnras, 415,
  3539, \dodoi{10.1111/j.1365-2966.2011.18968.x}

\bibitem[{{Viani} {et~al.}(2018){Viani}, {Basu}, {Joel Ong J.}, {Bonaca}, \&
  {Chaplin}}]{viani18}
{Viani}, L.~S., {Basu}, S., {Joel Ong J.}, M., {Bonaca}, A., \& {Chaplin},
  W.~J. 2018, \apj, 858, 28, \dodoi{10.3847/1538-4357/aab7eb}

\bibitem[{{Villaver} {et~al.}(2014){Villaver}, {Livio}, {Mustill}, \&
  {Siess}}]{viletal2014}
{Villaver}, E., {Livio}, M., {Mustill}, A.~J., \& {Siess}, L. 2014, \apj, 794,
  3, \dodoi{10.1088/0004-637X/794/1/3}

\bibitem[{{Vrard} {et~al.}(2015){Vrard}, {Mosser}, {Barban}, {Belkacem},
  {Elsworth}, {Kallinger}, {Hekker}, {Samadi}, \& {Beck}}]{vrard15}
{Vrard}, M., {Mosser}, B., {Barban}, C., {et~al.} 2015, \aap, 579, A84,
  \dodoi{10.1051/0004-6361/201425064}

\bibitem[{{Wittenmyer} {et~al.}(2017){Wittenmyer}, {Jones}, {Zhao}, {Marshall},
  {Butler}, {Tinney}, {Wang}, \& {Johnson}}]{witt17}
{Wittenmyer}, R.~A., {Jones}, M.~I., {Zhao}, J., {et~al.} 2017, \aj, 153, 51,
  \dodoi{10.3847/1538-3881/153/2/51}

\bibitem[{{Wittenmyer} {et~al.}(2016){Wittenmyer}, {Liu}, {Wang}, {Casagrand
  e}, {Johnson}, \& {Tinney}}]{witt16}
{Wittenmyer}, R.~A., {Liu}, F., {Wang}, L., {et~al.} 2016, \aj, 152, 19,
  \dodoi{10.3847/0004-6256/152/1/19}

\bibitem[{{Wu} \& {Li}(2016)}]{wu16}
{Wu}, T., \& {Li}, Y. 2016, \apjl, 818, L13,
  \dodoi{10.3847/2041-8205/818/1/L13}

\bibitem[{{Wu} \& {Li}(2017)}]{wu17}
---. 2017, \apj, 846, 41, \dodoi{10.3847/1538-4357/aa8361}

\bibitem[{{Wu} \& {Li}(2019)}]{wu19}
---. 2019, \apj, 881, 86, \dodoi{10.3847/1538-4357/ab2ad8}

\bibitem[{{Y{\i}ld{\i}z} {et~al.}(2019){Y{\i}ld{\i}z}, {{\c{C}}elik Orhan}, \&
  {Kayhan}}]{yildiz19}
{Y{\i}ld{\i}z}, M., {{\c{C}}elik Orhan}, Z., \& {Kayhan}, C. 2019, \mnras, 489,
  1753, \dodoi{10.1093/mnras/stz2223}

\bibitem[{{Zahn}(1977)}]{zahn1977}
{Zahn}, J.~P. 1977, \aap, 500, 121

\bibitem[{{Zhang} {et~al.}(2018){Zhang}, {Wu}, \& {Li}}]{zhang18}
{Zhang}, X., {Wu}, T., \& {Li}, Y. 2018, \apj, 855, 16,
  \dodoi{10.3847/1538-4357/aaaabb}

\end{thebibliography}
\bibliographystyle{aasjournal}

\clearpage

\begin{table}
\begin{center}
\caption{Stellar Parameters for HD~222076}
\label{tb:starpar}
\renewcommand{\tabcolsep}{0mm}
\begin{tabular}{l c c}
\noalign{\smallskip}
\tableline\tableline
\noalign{\smallskip}
\textbf{Parameter} & \textbf{Value} & \textbf{References} \\
\noalign{\smallskip}
\tableline
\noalign{\smallskip}
\multicolumn{3}{c}{Basic Properties} \\
\noalign{\smallskip}
\hline
\noalign{\smallskip}
TIC & 325178933 & 1 \\
\textit{Hipparcos} ID & 116630 & 2 \\
\textit{TESS} Mag. & 6.59 & 1 \\
Sp.~Type & K1 III & 3 \\
\noalign{\smallskip}
\hline
\noalign{\smallskip}
\multicolumn{3}{c}{Spectroscopy} \\
\noalign{\smallskip}
\hline
\noalign{\smallskip}
$T_{\rm eff}$ (K) & $4806 \pm 100$ & 4 \\
& $4900 \pm 100$ & 5\\
& $4834 \pm 59$ & 6\\
$[{\rm Fe}/{\rm H}]$ (dex) & $0.05 \pm 0.10$ & 4 \\
& $0.16 \pm 0.14$ & 5 \\
& $0.16 \pm 0.03$ & 6\\
$\log g$ (cgs) & $3.31 \pm 0.15$ & 4 \\
& $3.18 \pm 0.2$ & 5\\
& $3.24 \pm 0.13$ & 6 \\
\noalign{\smallskip}
\hline
\noalign{\smallskip}
\multicolumn{3}{c}{SED \& \textit{Gaia} DR2 Parallax} \\
\noalign{\smallskip}
\hline
\noalign{\smallskip}
$A_V$ & $0.08 \pm 0.02$ & 7 \\
$F_{\rm bol}$ (${\rm erg\,s^{-1}\,cm^{-2}}$) & $(3.57 \pm 0.13) \times 10^{-8}$ & 7 \\
$R_\star$ ($R_\sun$) & $4.38 \pm 0.20$ & 7 \\
$M_\star$ ($\msun$) & $1.38 \pm 0.14$\tablenotemark{a} & 7 \\
$\rho_\star$ (gcc) & $0.023 \pm 0.004$  & 7 \\
$L_\star$ ($L_\sun$) & $9.14 \pm 0.33$ & 7 \\
$\pi$ (mas) & $11.024 \pm 0.022$\tablenotemark{b} & 8 \\
\noalign{\smallskip}
\hline
\noalign{\smallskip}
\multicolumn{3}{c}{Asteroseismology} \\
\noalign{\smallskip}
\hline
\noalign{\smallskip}
$\Delta\nu$ ($\mu$Hz) & $15.60 \pm 0.13$ & 7 \\
$\nu_{\rm max}$ ($\mu$Hz) & $203.0 \pm 3.6$ & 7 \\
$\Dpi_{obs}$ (s) & $45.5 \pm 10.7$ & 7 \\
$M_\star$ ($\msun$) & $1.12 \pm 0.12$ & 7 \\
$R_\star$ ($R_\sun$) & $4.34 \pm 0.21$  & 7 \\
$\rho_\star$ (gcc) & $0.0194 \pm 0.0011$  & 7 \\
$\log g$ (cgs) & $3.214 \pm 0.053$  & 7 \\
$t$ (Gyr) & $7.4 \pm 2.7$  & 7 \\
\noalign{\smallskip}
\hline
\noalign{\smallskip}
\end{tabular}
\end{center}
\tablenotetext{a}{\scriptsize Based on extrapolated relations of \citet{torres10}, should be regarded with caution (cf. Section~\ref{sc:sed}) .}
\tablenotetext{b}{\scriptsize Adjusted for the systematic offset of \citet{stasstorres18}.}
\tablerefs{\scriptsize (1) \citet{stass18b}, (2) \citet{HIP}, (3) \citet{houk75}, (4) \citet{witt16}, (5) \citet{jones11}, (6) \citet{sousa18}, (7) this work, (8) \citet{GaiaDR2}.}
\end{table}

\begin{table}[!t]
\begin{center}
\caption{Extracted Oscillation Frequencies and Mode Identification for HD~222076}\label{tb:frequency}
\begin{tabular}{lccccc}
\hline
$\ell$  &$n$ & $n_{\rm p}$ & $n_{\rm g}$ &  $\nu$       &  $\sigma_{\nu}$ \\
          & & & & ($\muHz$) &    ($\muHz$)  \\
\hline
0   &9&9&\nodata& 161.06  &  0.03 \\
2   &-102&9&-111& 174.74  &  0.02 \\
0   &10&10&\nodata& 176.71  &  0.03 \\
1   &-51&10&-61& 184.81  &  0.07 \\
1   &-50&10&-60& 186.62  &  0.03 \\
2   &-92&10&-102& 190.31  &  0.02 \\
0   &11&11&\nodata& 192.32  &  0.01 \\
3   &-131&10&-141& 195.64  &  0.02 \\
1   &-46&11&-57& 198.11  &  0.02 \\
1   &-45&11&-56& 200.29  &  0.01 \\
1   &-44&11&-55& 202.34  &  0.03 \\
2   &-84&11&-95& 205.36  &  0.02 \\
2   &-83&11&-94& 205.92  &  0.03 \\ 
0   &12&12&\nodata& 207.75  &  0.02 \\
1   &-40&12&-52& 215.65  &  0.03 \\
1   &-39&12&-51& 217.50  &  0.02 \\
2   &-76&12&-88& 221.42  &  0.01 \\
0   &13&13&\nodata& 223.42  &  0.02 \\
3   &-109&12&-121& 226.86  &  0.03 \\
1   &-36&13&-49& 229.69  &  0.02 \\
1   &-35&13&-48& 231.26  &  0.02 \\
2   &-69&13&-82& 237.34  &  0.02 \\
0   &14&14&\nodata& 239.21  &  0.03 \\
1   &-31&14&-45& 246.95  &  0.03 \\
\hline
\end{tabular}
\end{center}
\tablecomments{Each mode is labeled according to its mode degree $\ell$, radial order $n$, radial order of p- and gravity-mode component $n_{\rm p}$ and $n_{\rm g}$ from the best-fitting model. $\nu$ is the mode cyclic frequency and $\sigma_\nu$ the uncertainty of $\nu$. Modes presented here are all with a height-to-background ratio larger than 4.}
\end{table}


\end{document}